\documentclass{elsart}
\usepackage{epsfig}
\DeclareGraphicsExtensions{.eps.gz,.eps,.ps,.ps.gz}

\def\Journal#1#2#3#4{{#1} {#2} (#4) #3}

\def\EPJC{{Eur. Phys. J.} C} 
\def\CPC{Comput. Phys. Commun.}

\def\NPB{{Nucl. Phys.} B}
 
\def\PLB{{Phys. Lett.} B} 
\def\PRL{Phys. Rev. Lett.} 
\def\PRD{{Phys. Rev.} D} 
\def\ZPC{{Z. Phys.} C} 
\def\PR{Phys. Rev.}

\def\PR{Phys. Rep.}

\newcommand{\Qsq}{\mbox{$Q^2~$}}
\newcommand{\qprimesq}{\mbox{$Q^{\prime\,2}~$}}
\newcommand{\xprime}{\mbox{$x^\prime~$}}
\newcommand{\iai}{I\overline{I}} 
\newcommand{\ai}{{\overline{I}}} 
\newcommand{\xpr}{{x^\prime}}
\newcommand{\ts}{\tilde{S}}
\newcommand{\dts}{D(\tilde{S})}
\newcommand{\ddts}{D\left( \ln \left( \frac{\dts}{\sqrt{\xi_\ast -2}}
                   \right)\right)}
\newcommand{\QISUDA}{y_{\rm Bj}\,e_t}

\newcommand{\QISUDC}{\sqrt{Sx_{\rm Bj}y_{\rm Bj}(1-y_{\rm Bj})}}
\newcommand{\XI}{\phi_q}
\newcommand{\ALR}{\frac{Q^{\prime 2}}{Sx^\prime z\, y_{\rm Bj}}}
\newcommand{\M}{P_t \left( \ALR \left(x_{\rm Bj} -x^\prime z 
                 + x_{\rm Bj} (1-y_{\rm Bj})\right)
                 -x_{\rm Bj}
       -\frac{m^2_{k}}{Sy_{\rm Bj}} \right)
       } 
\newcommand{\N}{e_t \frac{Q^{\prime 2}}{S x^\prime z}}
\newcommand{\Dtwo}{S x_{\rm Bj}y_{\rm Bj} \ALR \left(\ALR -1\right) 
-Q^{\prime 2} \left( \ALR -1\right) - \ALR m^2_{k}}
\newcommand{\D}{\sqrt{\Dtwo}}
\newcommand{\PHI}{\phi_{q^\prime}}
\newcommand{\Op}{2\, D\, \left( \cos\PHI\cos\XI+\sin\PHI\sin\XI\right)\QISUDC 
       \frac{P_t}{S y_{\rm Bj}}} 
\newcommand{\rb}[1]{\raisebox{1.5ex}[-1.5ex]{#1}}
  \newenvironment{nlist}[1]%
  {\begin{list}{}{\settowidth{\labelwidth}{#1}%
  \setlength{\leftmargin}{\labelwidth}%
  \addtolength{\leftmargin}{\labelsep}%
  \setlength{\itemsep}{0pt plus 1pt}
  \setlength{\parsep}{0pt plus 1pt}
  \setlength{\topsep}{0pt plus 1pt}
  \setlength{\partopsep}{0pt plus 1pt}
  \setlength{\parskip}{2mm plus 1mm minus 1mm}
  }}%
  {\end{list}}
\begin{document}

\begin{frontmatter}

\title{
{\rm\normalsize\rightline{DESY 99-180}\rightline{hep-ph/9911516}}
\vskip 1cm 
QCDINS 2.0 -- A Monte Carlo generator for 
instanton-induced processes in deep-inelastic 
scattering\thanksref{www}}

\thanks[www]{Information and code via WWW URL: 
www.desy.de/\~{}t00fri/qcdins/qcdins.html}

\author[DESY]{A. Ringwald\thanksref{emring}} and
\author[DESY]{F. Schrempp\thanksref{emschr}}

\address[DESY]{Deutsches Elektronen-Synchrotron DESY, 
D-22603 Hamburg, Germany}

\thanks[emring]{E-mail: ringwald@mail.desy.de}
\thanks[emschr]{E-mail: t00fri@mail.desy.de}

\begin{abstract}
We describe a Monte Carlo event generator for the simulation of QCD-instanton 
induced processes in deep-inelastic scattering (HERA). 
The QCDINS package is designed as an ``add-on'' hard process generator 
interfaced to the general hadronic event simulation package HERWIG. 
It incorporates the theoretically predicted production rate for
instanton-induced events as well as  
the essential characteristics that have been derived 
theoretically for the partonic final state of instanton-induced processes:
notably, the flavour democratic and isotropic production of the
partonic final state, energy 
weight factors different for gluons and quarks, and a high average
multiplicity 
$\mathcal{O}(10)$ 
of produced partons with a 
Poisson distribution of the gluon multiplicity.  
While the subsequent perturbative evolution of the generated partons
is always handled by the HERWIG package, the final hadronization step may
optionally be performed also by means of the general hadronic event simulation 
package JETSET.
\end{abstract}


\end{frontmatter}

\newpage

{\bf PROGRAM SUMMARY} \\ 

\begin{small}

\noindent {\em Title of program:\/} QCDINS 2.0 \\[10pt]
{\em Catalogue identifier:\/} \\[10pt]
{\em Program obtainable from:\/} 
http://www.desy.de/\~{}t00fri/qcdins/qcdins.html 
\\[10pt]
{\em Computer for which the program is designed and others on which it
has been tested:\/} Any computer with a FORTRAN 77 compiler\\[10pt]
{\em Operating systems 
under which the program has been
tested:\/} Linux 2.0.X; HP-UX 10.2\\[10pt]
{\em Programming language used:\/} FORTRAN 77\\[10pt]
{\em Memory required to execute with typical data:\/} Size of executable
program is approximately 2.6 MB. The size of the QCDINS library itself 
is about 200 KB; the required routines 
from the HERWIG and JETSET libraries constitute the dominant portion
of the needed memory. 
\\[10pt]
{\em No. of processors used:} 1 \\[10pt] 
{\em Has the code been vectorised or parallelized?:} no \\[10pt]
{\em No. of bytes in distributed program, including test data,
etc.:\/} 1071106 \\[10pt]
{\em Distribution format:\/} ASCII\\[10pt]
{{\em CPC Program Library subprograms used:\/}} 
HERWIG [1] version 5.9; JETSET 7.4 [2] 
\\[10pt]
{\em Nature of physical problem\/}\\
Instantons are a basic aspect of Quantum
Chromodynamics. Being non-perturbative fluctuations of the gauge
fields, they induce hard processes absent in conventional perturbation 
theory. Deep-inelastic lepton-nucleon scattering at HERA offers a
unique possible discovery window for such
processes induced by QCD-instantons through their characteristic
final-state signature and a sizable rate, calculable within
instanton-perturbation theory.  
An experimental discovery of such a novel, non-perturbative
manifestation of non-abelian gauge theories would be of fundamental
significance.  
However, instanton-induced events are expected to make up only a small fraction of all
deep-inelastic events. Therefore, a detailed knowledge of the
resulting hadronic final state, along with a multi-observable analysis of
experimental data by means of Monte Carlo techniques, is necessary.

{\em Method of solution\/}\\ 
The QCDINS package is designed as an ``add-on'' hard process generator 
interfaced to the general hadronic event simulation package HERWIG. 
It incorporates the theoretically predicted production rate for
instanton-induced events as well as  
the essential characteristics that have been derived 
theoretically for the partonic final state of instanton-induced processes:
notably, the flavour democratic and isotropic production of the
partonic final state, energy 
weight factors different for gluons and quarks, and a high average
multiplicity 
$\mathcal{O}(10)$
of produced partons with a 
Poisson distribution of the gluon multiplicity.  
While the subsequent perturbative evolution of the generated partons
is always handled by the HERWIG package, the final hadronization step may
optionally be performed also by means of the general hadronic event simulation 
package JETSET.

{\em Restrictions on the complexity of the problem\/}\\
The default values of the implemented kinematical cuts represent the
state of the art limits for the reliability of the generated instanton-induced
event rate and event topology. 

{\em Typical running time\/}\\
10 - 100 events per second for a PC with Pentium CPU, depending on its 
clock frequency. On a HP 9000/735 (99 MHz) workstation, 6 events per
second are generated.

{\em Unusual features of the program\/}\\
none

{\em References\/}\\
{[1]} G. Marchesini et al., Comput. Phys. Commun. {67} (1992) {465}.\\
{[2]} T. Sj{\"o}strand, Comput. Phys. Commun. {82} (1994) {74}.

\end{small}

\newpage

{\bf LONG WRITE-UP} \\

\section{Introduction}
The ground state (``vacuum'') of non-abelian gauge theories like QCD
is known to be very rich. It includes topologically non-trivial
fluctuations of the  gauge fields, carrying an integer topological
charge $Q$. The simplest building blocks of topological structure in the 
vacuum are~\cite{bpst,th} {\em instantons} with $Q=+1$ and 
{\em anti-instantons} with $Q=-1$. Instantons represent 
gluon field configurations that are localized (``instantaneous'') in
Euclidean time and space. 
While they are believed to play an important role in various
long-distance aspects of QCD, there are also important 
short-distance implications. In QCD with $n_f$ (light) flavours,
instantons induce hard processes violating {\em chirality} in
accord~\cite{th} with the selection rule  
$\Delta$ {\em chirality} $=2\,n_f\,Q$, due to the general chiral
anomaly. While
in ordinary perturbative QCD ($Q=0$) these
processes are forbidden, their experimental discovery would clearly be of basic
significance.   The deep-inelastic scattering regime is strongly 
favoured in this respect, since hard instanton-induced processes are both
calculable~\cite{mrs1,rs-pl} within instanton-pertubation theory and have good
prospects for experimental detection at HERA~\cite{rs-pl,rs,grs,cgrs}.

QCDINS~\cite{grs} is a Monte Carlo package for simulating 
QCD-instanton induced scattering processes in deep-inelastic
scattering (HERA). It is designed as an  
``add-on'' hard process generator interfaced by default to the Monte
Carlo generator HERWIG~\cite{herwig}. 
It incorporates the theoretically predicted production rate for
instanton-induced events as well as  
the essential characteristics that have been derived 
theoretically for the partonic final state of instanton-induced processes:
notably, the flavour democratic~\cite{th} and
isotropic~\cite{rs,dis97} production of the final 
state partons, energy
weight factors different for gluons and quarks~\cite{mrs1}, and a high average
multiplicity $2n_f+\mathcal{O}(1/\alpha_s)$ of produced partons with a 
(approximate) Poisson distribution of the gluon 
multiplicity~\cite{rs,dis97,dis97-phen}.  

Earlier versions of QCDINS have been used already to establish first
experimental bounds on the rate of instanton-induced events at 
HERA~\cite{k-bound,mult-bound,ck} and to develop instanton search 
strategies~\cite{cgrs}.

In the present report a comprehensive description is given of the theoretical 
framework built into the program (Section~\ref{theorie}) as well as of
the various program components (Section~\ref{QCDINS}) and of their
usage (Section~\ref{usage}).

\section{\label{theorie}Instanton-induced events in deep-inelastic scattering}

Let us briefly summarize in this section the underlying physics picture,
some relevant formulae and the main stages involved in
QCDINS to generate the complete instanton-induced partonic final
state. The remaining formulae may be found under the corresponding
descriptions of QCDINS routines in Section~\ref{QCDINS}.  

\begin{figure}[h]
\vspace{3ex}
\begin{tabular}{ll}
\mbox{
 \epsfig{file=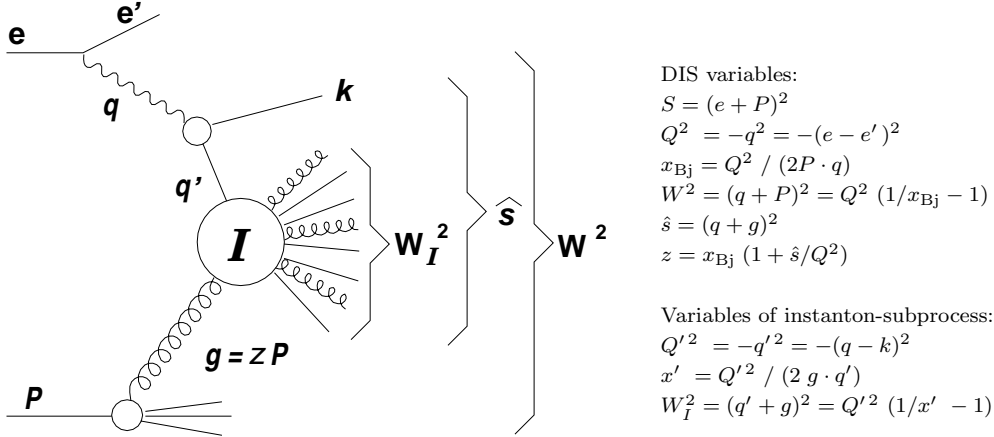,width=8cm,%
 bbllx=0pt,bblly=0pt,bburx=318pt,bbury=232,clip=}}
&
{\scriptsize
\renewcommand{\arraystretch}{1.2}
\begin{tabular}{l}
 \vspace{-5.5cm} \\
 DIS variables: \\
   $S=(e+P)^2$\\
   $\Qsq = - q^2 = -(e-e^\prime\,)^2$\\ 
   $ x_{\rm Bj} = \Qsq / \; (2 P \cdot q) $ \\
   $ W^2 =(q+P)^2 = \Qsq (1/x_{\rm Bj}-1)$ \\
   $ \hat{s} = (q+g)^2$ \\
   $ z = x_{\rm Bj} \;(1+\hat{s}/Q^2)$ \\ \\
 Variables of instanton-subprocess: \\
 $\qprimesq=- q^{\prime\,2} = - (q-k)^2 $ \\
 $\xprime= \qprimesq / \;(2 \; g \cdot q^\prime ) $ \\
 $W_I^2= (q^\prime+g)^2 = \qprimesq ( 1/\xprime -1)$\\
\end{tabular}}
\end{tabular}
\vspace{0.3cm}
   \caption[]
     {Structure and kinematic variables of the dominant
     instanton-induced process in deep-in\-elastic scattering.} 
   \label{kin-var}
\end{figure}
 
In deep-inelastic $e^\pm P$ scattering, instanton-induced events
are predominantly associated~\cite{rs-pl} with a process structure as
sketched in Fig.~\ref{kin-var}: A photon splitting into a
$q\overline{q}$-pair fuses with a gluon from the proton in the background of an
instanton ($I$) or an anti-instanton ($\ai$). For each (light)
flavour, $q=\{d,u,s,\ldots\}$, a violation of {\em chirality} is induced, 
\begin{equation}
\label{anomaly}
\Delta\, \mbox{\em chirality}\, (q) \equiv 
\Delta  \left[\#(q_R+\overline{q}_R) -\#(q_L+\overline{q}_L)\right]=\pm\, 2
\ \mbox{\ for an} 
\left\{\begin{array}{l}I\\\ai 
\end{array} 
\right.
\,,
\end{equation}
in agreement with the general chiral anomaly
relation~\cite{th}. 
Correspondingly, the partonic final state exhibits ``flavour
democracy'', i.\,e. $q_R\overline{q}_R$ ($q_L\overline{q}_L$) pairs of
{\em all} $n_f$ light flavours occur precisely once in case of an
instanton (anti-instanton),
\begin{equation}
\label{flavour-democr}
\gamma^\ast + g\ \stackrel{I\,(\overline{I})}{\Rightarrow}
\sum_{q=d,u,s,\ldots} \left[ q_{R\,(L)}+\overline{q}_{R\,(L)}\right]
+n_g\,g
\,.
\end{equation}
 As illustrated in Fig.~\ref{kin-var},
one of those partons acts as
a current-quark (jet) $k$, whereas the other   
$2n_f-1$  (anti-)quarks and some number $n_g$ of gluons are directly
emitted from  the instanton (anti-instanton) ``blob''. Instanton-induced 
processes initiated by a quark from the proton are 
suppressed by a factor of $\alpha_s^2$ with respect to the gluon initiated
process~\cite{rs-pl}. This fact, together with 
the high gluon density in the relevant kinematical domain at HERA, 
justifies to neglect  quark initiated processes.

In instanton-perturbation theory, the dominant
instanton-induced contribution to the inclusive $eP$ cross 
section\footnote{A sum over instanton-induced and anti-instanton 
induced processes is always implied by the superscript $(I)$ 
at cross sections.},
subject to appropriate kinematical restrictions and (theoretical) fiducial 
cuts, has a convolution-like structure~\cite{rs-pl},
\begin{eqnarray}
          \nonumber 
          \sigma_{eP}^{(I)}({\rm cuts}) &\simeq & 
          \frac{2\,\pi\,\alpha^2}{S}\,\sum\limits_{q^\prime =
          d,u,s,\ldots ;\,
          \overline{d},\overline{u},\overline{s},\ldots } e_{q^\prime}^2
          \ \int\limits_{Q^{\prime 2}_{\rm min}}^{Q^{\prime 2}_{\rm max}} 
          dQ^{\prime 2}\,
          \int\limits_{x^\prime_{\rm min}}^{x^\prime_{\rm max}}
          \frac{dx^\prime}{x^\prime}\,
          \,\frac{\sigma^{(I)}_{q^\prime g}(x^\prime ,Q^{\prime 2})}{x^\prime}
          \\[2ex]  \label{evwgt}
          && \times
          \int\limits_{{\rm max}\left(\frac{Q^{\prime 2}}{Sx^\prime 
          y_{\rm Bj\, max}},\frac{x_{\rm Bj\,min}}{x^\prime}
          \right)}^{z_{\rm max}}
          \frac{dz}{z}\,f_g (z )
          \int\limits_{x_{\rm Bj\,min}}^{x^\prime z -
          \frac{m_{k}^2}{S} 
          \frac{1}{y_{\rm Bj\,max}-\frac{Q^{\prime 2}}{Sx^\prime z}}} 
          \frac{dx_{\rm Bj}}{x_{\rm Bj}}\, 
          \\[2ex] \nonumber && \times
          \int\limits_{{\rm max}\left( 
          \frac{Q^{\prime 2}}{Sx^\prime z}+\frac{m_{k}^2}{S}
          \frac{1}{x^\prime z -x_{\rm Bj}},y_{\rm Bj\,min}
          \right)}^{y_{\rm Bj\,max}}
          \frac{dy_{\rm Bj}}{y_{\rm Bj}}\,
          \theta (S x_{\rm Bj} y_{\rm Bj} - Q^2_{\rm min})\,
          \\[2ex] \nonumber && \times
          \frac{1+(1-y_{\rm Bj})^2}{y_{\rm Bj}}\  
          P_{q^\prime}^{{ (I)}}\ 
          .
\end{eqnarray}  
It involves integrations over the gluon density $f_g(z)$, the virtual 
photon flux and the flux of virtual (anti-)quarks $q^\prime$ in the 
instanton-background~\cite{rs-pl,mrs2},
\begin{equation}
\label{flux}
P_{q^\prime}^{{ (I)}} = \frac{3}{16\,\pi^3}\,\frac{x_{\rm Bj}}{z\, x^\prime}
\left(1+\frac{z}{x_{\rm Bj}}
-\frac{1}{x^{\prime}}-\frac{Q^{\prime 2}}{S x_{\rm Bj}y_{\rm Bj}}\right)
\, .
\end{equation}
All relevant kinematical variables 
in Eq.~(\ref{evwgt}) are defined in Fig.~\ref{kin-var}, 
and $e_{q^\prime}^2$ denotes the electric charge 
squared of the virtual (anti-)quark $q^\prime$ in units of the electric charge
squared $e^2=4\pi\alpha$. 

\begin{figure} [ht]
\centering
\mbox{
\epsfig{file=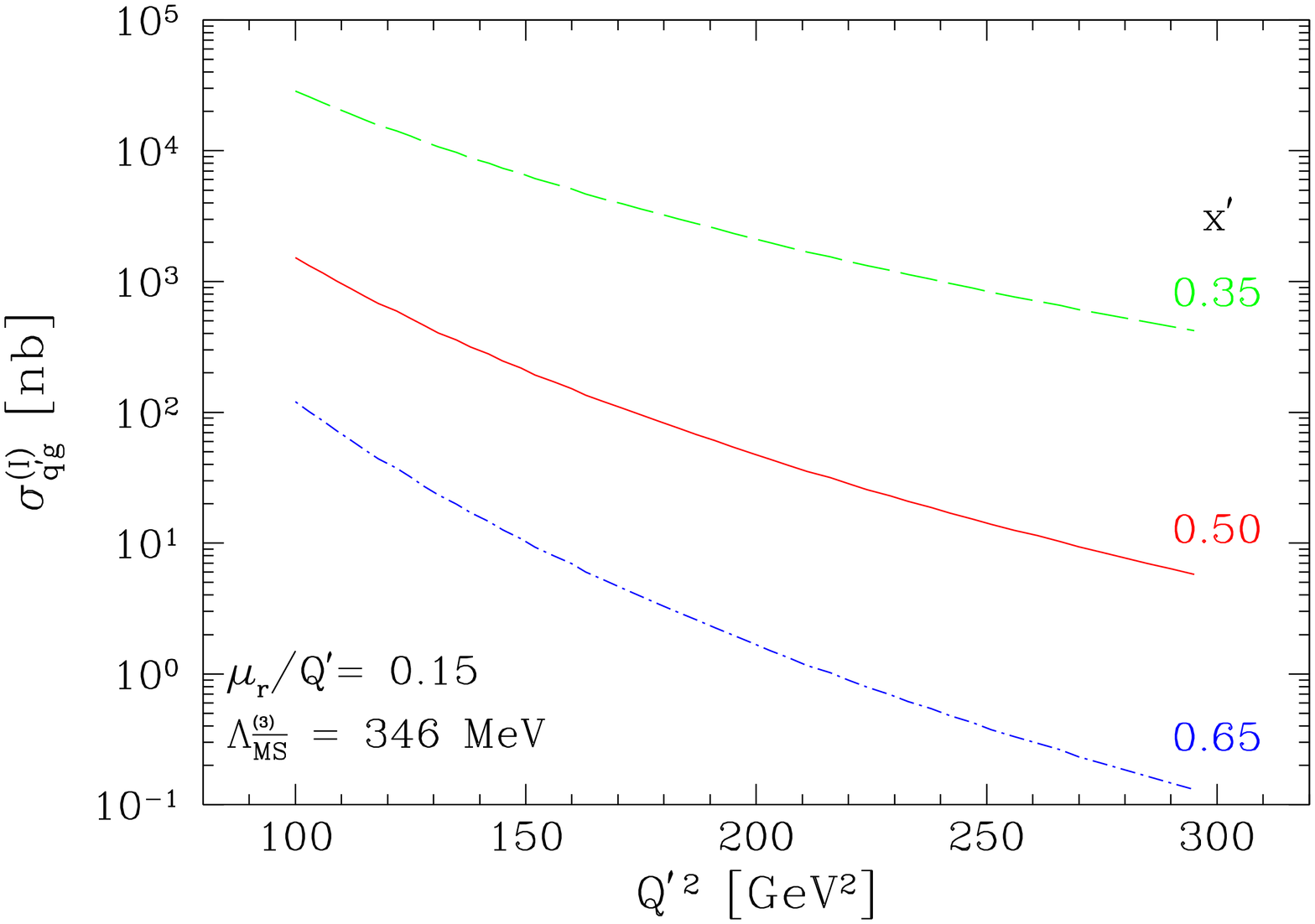,angle=-90,width=6.7cm}
\epsfig{file=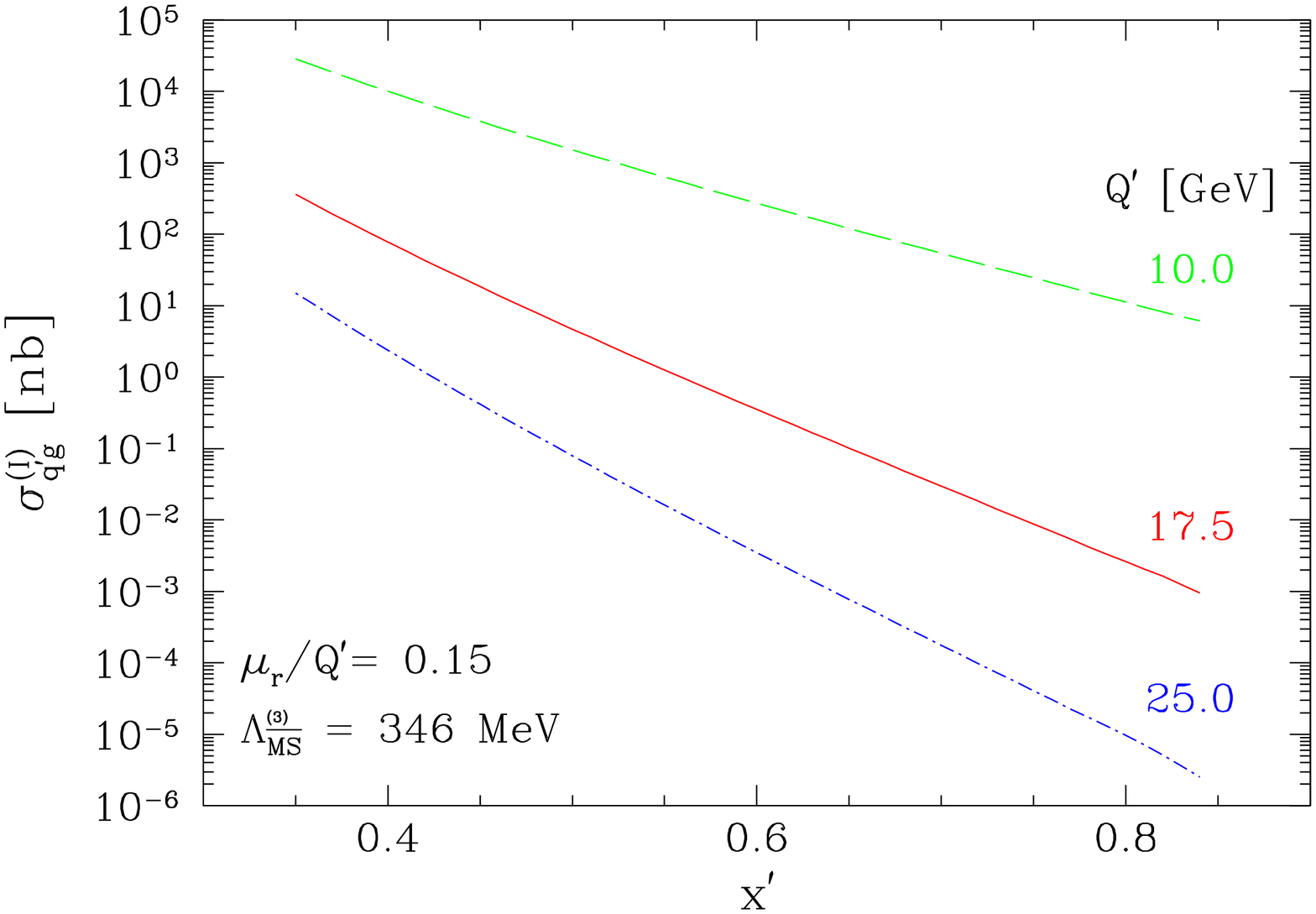,angle=-90,width=6.7cm}
     }
  \caption[]
   {Instanton-subprocess cross section~(\ref{qgcross-vxi}) from
   Ref.~\cite{rs-pl} for $n_f=3$ and $\Lambda^{(3)}_{\overline{\rm MS}}$ 
from Eq.~(\ref{lambda}). It  exhibits a strong dependence on the corresponding
    Bjorken variables $Q^{\prime\,2}$ and $x^\prime$, respectively.}
  \label{isorho}
\end{figure}

In Eq.~(\ref{evwgt}), $\sigma_{q^\prime g}^{(I)}$ denotes the
instanton-induced  total  cross section  of the $q^\prime
g$-subprocess (c.\,f. Fig.~\ref{kin-var}) and contains the essential
instanton dynamics. Its analytical form~\cite{rs-pl} used in QCDINS
may be found in Eq.~(\ref{qgcross-vxi}).  
As illustrated in Fig.~\ref{isorho}, $\sigma_{q^\prime g}^{(I)}$ is
very steeply growing  for decreasing values of $Q^{\prime 2}$ and
$x^\prime$, respectively. Eventually, our theoretical predictions
based on instanton-perturbation theory~\cite{rs-pl} will 
cease to hold. Therefore, the following cuts inferred from a
high-quality lattice simulation of QCD~\cite{rs-lat}
are implemented by default in QCDINS 2.0 (Table~\ref{floating}),
\begin{equation} 
\label{fiducial}
Q^{\prime}\geq 
Q^{\prime}_{\rm min} = 30.8\,\Lambda^{(n_f)}_{\overline{\rm MS}} \, ; 
\hspace{6ex}
x^\prime \geq x^\prime_{\rm min} = 0.35 .
\end{equation}
A further cut on the photon virtuality,
\begin{equation}
\label{fiducialQ}
  Q\equiv \sqrt{S x_{\rm Bj} y_{\rm Bj}}\geq Q_{\rm
min}=Q^{\prime}_{\rm min}  , 
\end{equation}
is applied in order to warrant sufficient  suppression of non-planar
contributions~\cite{mrs1}, which are hard to calculate and may spoil the  
validity of Eq.~(\ref{evwgt}).

The cross section $\sigma_{q^\prime g}^{(I)}$ exhibits a rather weak
residual dependence on the renormalization scale $\mu_r$. As an
``optimal'' choice, the value $\mu_r = 0.15\,Q^\prime$,
corresponding to the minimum~\cite{rs-pl}, $\partial
\sigma_{q^\prime\,g}/\partial\mu_r = 0$, is taken 
by default (Table~\ref{floating}).

However, $\sigma_{q^\prime g}^{(I)}$
depends strongly on the QCD scale
$\Lambda^{(n_f)}_{\overline{\rm MS}}$. 
Since, strictly speaking, the underlying theoretical framework
refers to massless quarks, the (default) number of flavours
is set to $n_f=3$ (Table~\ref{floating}). The respective value 
\begin{equation}
\Lambda^{(3)}_{\overline{\rm MS}}=0.346\,{{+0.031}\atop{-0.029}} 
\ {\rm GeV},
\label{lambda}
\end{equation}
is obtained
by a standard 3-loop perturbative flavour reduction (Eq.~(9.7)
of Ref.~\cite{pdg}) from the 1998 world-average
of the running QCD coupling at the Z-boson mass~\cite{pdg},  
\begin{equation}
\alpha_s(M_Z)=0.119\,\pm\, 0.002\ \ 
\stackrel{\rm 3-loop}{\Leftrightarrow}\ \  
\Lambda^{(5)}_{\overline{\rm MS}} = 
0.219\,{{+0.025}\atop {-0.023}}\mbox{\ GeV}. 
\end{equation}
The central values of these parameters are taken as default in QCDINS
2.0 (Table~\ref{floating}). This upgrade of
    $\Lambda^{(n_f)}_{\overline{\rm MS}}$, 
together with the modified cuts (\ref{fiducial}) and 
(\ref{fiducialQ}),  represents 
an improved understanding of the input parameters and  a considerable
reduction of  
uncertainties, as compared to  the original publication~\cite{rs-pl}
and earlier  
versions of QCDINS. Note that it also implies a significant change in
the predicted rate. 

Next, let us summarize the various stages of event generation by means 
of QCDINS.

In a first stage, the various Bjorken variables $Q^{\prime
2},x^\prime,z,x_{\rm Bj}, y_{\rm Bj}$    
of the instanton-induced process (c.\,f. Fig.~\ref{kin-var}) are
generated, with a distribution according to the normalized differential cross 
section from Eq.~(\ref{evwgt}),
\begin{equation}
\label{diffcross}
\frac{1}{\sigma_{eP}^{(I)}}\ 
\frac{d^5\sigma_{eP}^{(I)}}
{dQ^{\prime 2}\,dx^\prime\, dz\, dx_{\rm Bj}\, dy_{\rm Bj}} 
\, .
\end{equation} 

In the second stage of the event generation, the 4-momenta $g,q,q^\prime,
k$ of the incoming gluon $g$, the virtual photon $q$, 
the virtual quark $q^\prime$ and the current quark 
$k$, respectively, are filled. Sudakov
decompositions of these momenta are used to incorporate various
constraints, e.g. on the momenta squared. The 4-momentum $e^\prime$
of the outgoing lepton is calculated subsequently. 

In the third stage, the partonic final state of the instanton-induced 
$q^\prime g$-subpro\-cess is generated in its centre-of-mass system (CMS) as 
follows.
The number $n_g$ of produced gluons is generated according to 
a Poisson distribution with mean $\langle n_g\rangle^{(I)}(\xpr,Q^\prime) \sim
1/\alpha_s\sim 3$ (for the cuts (\ref{fiducial})), as calculated theoretically 
(Eq.~(\ref{ng})) in
instanton-perturbation theory~\cite{rs,dis97,dis97-phen}. Next,
$n_f (=3)$ $[\,q\,\ldots\,\overline{q}\,]$-``strings'' of partons are 
set up, each beginning with a quark, 
followed by a random number $\leq n_g+1$ of gluons and ending with an 
anti-quark of randomly chosen flavour. There are $n_g+1$ gluons in total and, 
due to the required flavour democracy~(\ref{flavour-democr}),
$q\overline{q}$-pairs of all  $n_f$ flavours occur precisely once.
A quark and a gluon among these $2n_f+n_g+1$ partons are (randomly) marked
as incoming.  

The momenta $p_i$ of the $n=2n_f-1+n_g$ outgoing partons are then
generated in the CMS of the instanton subprocess, according to the  
energy-weighted phase-space 
\begin{eqnarray}
\label{leadingorder}
\lefteqn{\int 
\prod_{i=1}^{2n_f-1}
\left\{ d^4p_i\,\delta^{(+)}\left( p_i^2-m_i^2\right) p^0_i\right\}
\prod_{k=1}^{n_g}
\left\{ d^4p_k\,\delta^{(+)}\left( p_k^2-m_g^2\right) p^{0\,2}_k\right\}}
\\ \nonumber &&\times
\, \delta^{(3)}\left( \sum_{i=1}^{2n_f-1} \vec{p}_i+
\sum_{k=1}^{n_g} \vec{p}_k \right)\,
\delta \left( W_I - \sum_{i=1}^{2n_f-1} p_i^0 - 
\sum_{k=1}^{n_g} p_k^0\right) . 
\end{eqnarray} 
These different energy weights for quarks and gluons~\cite{mrs1},
along with the angular isotropy~\cite{rs,dis97}, are characteristic
features of the leading-order partonic final state (after averaging
over colour). 

Next, the colour and flavour connections of the partons are set up. The
colour flow is obtained  simply by 
connecting the colour lines of adjacent partons within each of the
above-mentioned $n_f$ $[\,q\,\ldots\,\overline{q}\,]$-``strings'' in  a
planar manner (consistent with the leading order $1/N_c$ expectation).  
This choice is inspired by the leading-order partonic final state (after 
averaging over colour)~\cite{mrs1}, but may well deserve further research.
The flavour flow is constructed by connecting the flavour lines of the 
quark at the beginning of a string with the flavour line of 
the anti-quark at the end of a string.

The hard subprocess generation ends by boosting
the momenta of the final state partons to the laboratory frame. 

While the subsequent perturbative evolution of the generated partons
is always handled 
by the HERWIG~\cite{herwig} package, the final hadronization step may
optionally\footnote{We thank H. Jung for his help~\cite{HJung} to interface 
QCDINS with JETSET.} be performed also by means of JETSET~\cite{jetset}.

\section{\label{QCDINS} The QCDINS package}

This section is devoted to a systematic description of the various
routines of the QCDINS package that is designed as an  
``add-on'' hard process generator, interfaced to the Monte
Carlo generator HERWIG~\cite{herwig}.   

This reference section is organized as follows:

While all subroutines and functions of the QCDINS package are
described in {\em alphabetical order} in Section~\ref{routines}, 
a {\em logical flow-chart} is provided in form of Tables~\ref{flow-chart}
and \ref{flow-qihgen} below. They should always be used as the main
guide through the description of the package.  
A routine listed in the n-th column and the m-th row of these tables
progressively 
calls all routines in the (n+1)-th column starting in the (m+1)-th row. 
All routines called at the level of the main hard process generator
{\bf QIHGEN} and below are documented in Table~\ref{flow-qihgen}.

A specific application requires the writing of a steering program by the
user (c.\,f. Section~\ref{usage} and Appendix A). It must contain the standard
HERWIG-initialization 
calls as well as the calls to various initialization routines for
QCDINS. The latter comprise essentially the four routines listed in
the second column of Table~\ref{flow-chart}. 
By calling the last one of these ({\bf QCLOOP}), the user starts    
the proper simulation which comprises the chain of internally called
QCDINS routines as  
documented in Tables~\ref{flow-chart} and \ref{flow-qihgen}. 
A specific and quite extensive example is provided with the QCDINS
distribution ('qtesthz.F') and may be found in the directory 'qcdtest'.
It also illustrates the use of the event analysis routine
{\bf HWANAL} called by HERWIG after each processed event.

\begin{table}
\begin{center}
{\scriptsize
\renewcommand{\arraystretch}{1.4}
\begin{tabular}{|l||c|c|c|c|c|}\hline
Initialize input para-&&&&&\\
meters (Tables~\ref{floating},~\ref{flags}) & \rb{\bf QIINIT} & & && \\ \hline
Flavour reduction: &&&&&\\
$\Lambda_{\overline{\rm MS}}^{(5)}\Rightarrow \Lambda_{\overline{\rm
MS}}^{(n_f<5)}$ && \rb{\bf LAMNF} &&& \\ \hline
Initialize ``particle" &&&&&\\ 
INST in event record &  & \rb{\bf QIINIH} &&& \\ \hline
Print input parame- &&&&&\\
ters and warnings  &  \rb{\bf QISTAT} &  &&&\\ \hline
Initialize JETSET &&&&&\\ 
common block data &\rb{\bf GJEINI} & &&&\\ \hline
Loop over desired &&&&&\\
number of events &  \rb{\bf QCLOOP} & &&&\\ \hline
Generate one event & &  {\bf QCDGEN} &&&\\ \hline
Assign hard process &&&&&\\ 
variables (HERWIG) & & &  \rb{\em HWEPRO} &&\\ \hline
No action (modified &&&&&\\ 
HERWIG routine)& & & &  \rb{\bf HWEGAM} &\\ \hline
Call main instanton &&&&&\\
process generator & & & &  \rb{\bf HVHBVI} &\\ \hline
Main (hard) instanton &&&&&\\
process generator & & & & &
\rule[-2.3ex]{0ex}{4ex}\rb{\framebox[16ex]{\bf QIHGEN}} \\ \hline 
Generate parton &&&&&\\ 
cascades (HERWIG) & & & \rb{\em  HWBGEN} & & \\ \hline 
Combine jets with &&&&& \\ 
correct kinematics & & & &   \rb{{\bf HWBJCO}} & \\ \hline
Convert HERWIG to &&&&&\\ 
JETSET block data & & &   \rb{\bf HERLUND} & & \\ \hline
JETSET event record &&&&&\\ 
to HEPEVT common & & &   \rb{\bf LUHEPC} & & \\ \hline  
\end{tabular}
}
\end{center}
\vspace{2ex}
\caption[]{\label{flow-chart} Flow-chart of QCDINS routines:
A routine in the n-th column and the m-th row progressively
calls all routines in the (n+1)-th column starting in the (m+1)-th row. 
All routines called at the level of the main hard process generator
{\bf QIHGEN} and below are listed in Table~\ref{flow-qihgen}.}
\end{table}

\begin{table}
\begin{center}
{\scriptsize
\begin{tabular}{|l||c|c|c|}\hline
Optional account of initial state &  &&\\ 
radiation from lepton & \rb{\bf EXFRAC} &&\\ \hline
Check kinematical boundaries & \bf QICALC &&\\ \hline
Generate identity  code of current &  &&\\ 
quark $k$ and virtual quark $q'$ &\rb{\bf QIHPAR}&&\\ \hline 
Generate $(Q', x')$ and associated weight & \bf QIHINS &&\\ \hline
Generate $X=Q'^2,x'$ as $dX/X^{(N+1)}$ & & \bf QIRDIS &\\ \hline
Calculate total cross section of & & &\\ 
instanton-induced subprocess $q'g \stackrel{(I)}{\Rightarrow}X$ & & \rb{\bf
Q2SIG} &\\ \hline  
Calculate $\iai$-valley action $S^{(\iai )}(\xi )$& & & \bf ACTION \\ \hline
Calculate fermionic overlap $\omega(\xi)$ & & & \bf OMEGA \\ \hline
Calculate Lambert W-function & & & \bf LAMBERTW \\ \hline
Calculate saddle-point value of  & & & \\  
conformally invariant $\iai$-distance
$\xi$ & & & \rb{\bf XI} \\ \hline 
\rule[-2.3ex]{0ex}{4ex}Calculate inverse running coupling
$\frac{4\pi}{\alpha_{\overline{\rm MS}}}$ & & &
\rule[0ex]{0ex}{4ex}{\bf XQS} \\ \hline   
Generate number $n_g$ of emitted gluons & \bf QIGMUL && \\ \hline
Calculate gluon multiplicity $\langle n_g \rangle^{(I)} $ 
& & \bf GMULT &\\ \hline
\rule[-2.3ex]{0ex}{4ex}Calculate inverse running coupling
$\frac{4\pi}{\alpha_{\overline{\rm MS}}}$ & & &
\rule[0ex]{0ex}{4ex}{\bf XQS} \\ \hline  
Calculate $\iai$-valley action $S^{(\iai )}(\xi )$& & & \bf ACTION \\ \hline
Calculate Lambert W-function & & & \bf LAMBERTW \\ \hline
Calculate flux of virtual quark $q'$ & \bf QISPLT &&\\ \hline
Calculate remaining weight & \bf QIPVWT &&\\ \hline
Calculate momentum of incoming gluon & \bf QIKPAR &&\\ \hline
Generate momentum of virtual photon & \bf QIKGAM &&\\ \hline
Generate momenta of virtual quark $q'$ &  &&\\
and current quark $k$ & \rb{\bf QIKGSP} &&\\ \hline
Generate partonic final state & \bf QISTID &&\\ \hline
Generate the partons of the instanton  &&&\\
subprocess in form of $[q\ldots\overline{q}]$-''strings''&& \rb{\bf QIGLST}
&\\ \hline 
Find the incoming partons && &\\
in the $[q\ldots\overline{q}]$-strings  && \rb{\bf QIGPAR} &\\ \hline
Assign masses of outgoing partons && \bf QIPLST &\\ \hline
Generate 4-momenta of outgoing partons && \bf QIPSGN &\\ \hline
Calculate relative energy weight &&&\\
of outgoing partons  &&& \rb{\bf QIPSWT} \\ \hline
Store 4-momenta of outgoing partons &&&\\
into PHEP common block of HERWIG && \rb{\bf QIPSTO} &\\ \hline
Colour/flavour connections for each string && \bf QICCON &\\ \hline
\end{tabular}
}
\end{center}
\vspace{2ex}
\caption[]{\label{flow-qihgen} All routines called at the level of the
main hard process generator {\bf QIHGEN} and below. As in
Table~\ref{flow-chart}, a routine in the n-th
column and the m-th row progressively calls all routines in the (n+1)-th 
column starting in the (m+1)-th row. The displayed sequence of calls 
corresponds to the default settings of control flags.}
\end{table}
\vfill\eject

\subsection{\label{routines} Subroutines and functions}

\noindent
SUBROUTINE {\bf ACTION}\,(XI4,\,S,\,DS,\,DDS)
\begin{nlist}{abcdefghikl}
\item[\em Purpose:]
Calculation of the $\iai$-action as well as
its 1st and 2nd derivatives,  as function of the conformally invariant
$\iai$-distance.\\
\item[\em Arguments:]\hfill

\begin{nlist}{abcdef}
\item[XI4:]   conformally invariant $\iai$-distance $\xi$.
\item[S:]   $\iai$-action $S^{(\iai )}(\xi )$, Eq.~(\ref{action}).
\item[DS:]   $dS^{(\iai )}/d\xi$,
\item[DDS:]   $d^2S^{(\iai )}/d\xi^2$
\end{nlist}
\item[\em Procedure:]
\end{nlist}
The action is calculated 
according to the exact valley form~\cite{optvalleyqcd,verb}, 
\begin{eqnarray}
\label{action}
S^{(\iai )}(\xi )   &=& 1-\frac{12}{f(\xi )} 
- \frac{96}{f(\xi )^2} +\frac{48}{f(\xi )^3}\left[ 3f(\xi )+8\right]
\ln\left[ \frac{1}{2\xi }\bigl( f(\xi ) +4\bigr)\right] \, ,
\\[2.4ex]
f(\xi )&=&\xi^2+\sqrt{\xi^2-4}\xi-4 \,  .     
\end{eqnarray}

\vspace{3ex}
\noindent
SUBROUTINE {\bf EXFRAC}\,(A) 
\begin{nlist}{abcdefghikl}
\item[\em Purpose:] 
Optional account of initial state radiation from the lepton.\\
\item[\em Arguments:]\hfill 
\begin{nlist}{ab}
\item[A:] 
\begin{nlist}{abcdef}
\item[= 1.0;] dummy rescaling factor of the incoming lepton 
momentum; double precision output variable.
\end{nlist}
\end{nlist}
\item[\em Remarks:]
The actual routine has to be provided by the user.
\end{nlist}

\noindent
SUBROUTINE {\bf GJEINI} 
\begin{nlist}{abcdefghikl}
\item[\em Purpose:]
Initialization  of the JETSET~\cite{jetset} parameter common blocks
LUDAT1, LUDAT2, LUDAT3, LUDAT4 and  LUDATR.\\ 
\item[\em Remarks:]{\bf GJEINI} has to be called by the user before
any other JETSET routine. {\bf GJEINI} is from Ref.~\cite{ggrind}. 
\end{nlist}

\vspace{15ex}

\noindent
FUNCTION\vspace{-2ex}

\noindent
\begin{flushright}
{\bf GMULT}\,(XPR,\,XI\_MIN,\,XI\_MAX,\,QLAM,\,KAP\-PA,\,NF,\,LOOPFL)
\end{flushright}
\begin{nlist}{abcdefghikl}
\item[\em Purpose:]
Calculation of the average gluon multiplicity
$\langle n_g\rangle^{(I)}$ depending on 
$x^\prime$, $Q^{\prime}/\Lambda^{(n_f)}_{\overline{\rm MS}}$,
$\mu_r/Q^\prime$, $n_f$ and loop-order. Here, $\mu_r$ and $n_f$
denote the renormalization scale and the number of light flavours,
respectively.\\  
\item[\em Arguments:]\hfill
\begin{nlist}{abcdefghi}
\item[XPRIME:] $x^\prime$
\item[XI\_MIN:]  $\xi_{\rm min}$; lower boundary of $\xi$ used for 
interpolation.
\item[XI\_MAX:]  $\xi_{\rm max}$; upper boundary of $\xi$ used for 
interpolation. 
\item[QLAM:] $Q^{\prime}/\Lambda^{(n_f)}_{\overline{\rm MS}}$
\item[KAPPA:] $\mu_r/Q^\prime$
\item[NF:] $n_f$; number of light flavours.
\item[LOOPFL:] 
\begin{nlist}{abcd}
\item[= 1:] 1-loop renormalization group (RG) invariance~\cite{rs-pl} along 
            with 1-loop form of $\alpha_s$.
\item[= 2:] 2-loop RG invariance~\cite{rs-pl} along with
            2-loop form of $\alpha_{\overline{\rm MS}}$.
\item[= 3:] (default) 2-loop RG invariance along with
            3-loop form of $\alpha_{\overline{\rm MS}}$.
\end{nlist}
\end{nlist}
\item[\em Procedure:]
\end{nlist}
From an analysis based on the  generalized
(Mueller~\cite{mueller-opt}) optical theorem for the $q^\prime g
\overline{g}$ forward scattering amplitude and the $\iai$-valley
method, one infers~\cite{dis97} the differential one-gluon inclusive 
$q^\prime g\stackrel{I}{\Rightarrow} g+X$ cross section,
normalized by the total cross section $\sigma_{q^\prime g}^{(I)}$. The mean 
gluon multiplicity~\cite{rs,dis97-phen} is then found by phase space
integration, 
\begin{equation}
\label{ng}
\langle n_g\rangle^{(I)}\left(x^\prime ,\frac{Q^\prime}
{\Lambda_{\overline{\rm MS}}^{(n_f)}},\frac{\mu_r}{Q^\prime},n_f\right)=
\frac{2\,\pi}{\alpha_{\overline{\rm MS}}(1/\rho^\ast )}\,
(\xi_\ast -2)\,\frac{dS^{(I\overline{I})}}{d\xi}(\xi_\ast ) .
\end{equation}
The function {\bf GMULT} calculates and returns the average gluon 
multiplicity~(\ref{ng}). 

The stars $(\ast )$ in Eq.~(\ref{ng}) denote the saddle point values
of the $\iai$ collective coordinates $\rho$, $\overline{\rho}$ and
$\xi$. Their computation proceeds as in the descriptions of the
functions {\bf Q2SIG} and {\bf XI}.  
The required values of $4\pi/\alpha_{\overline{\rm MS}}$ are calculated
and returned by the function {\bf XQS}. 
The $\iai$-action $S^{(\iai )}(\xi )$ and its $\xi$-derivatives are
provided by the subroutine {\bf ACTION}.  

\vspace{10ex}
\noindent
SUBROUTINE {\bf HERLUND} 
\begin{nlist}{abcdefghikl}
\item[\em Purpose:]
Conversion of the HERWIG~\cite{herwig}
event record in the HEPEVT common block to the respective
JETSET~\cite{jetset} common block.\\
\item[\em Remarks:]{\bf HERLUND} is a
modified~\cite{HJung} copy from the JETSET subroutine
LUHEPC.  
\end{nlist}

\noindent
SUBROUTINE {\bf HVHBVI} 
\begin{nlist}{abcdefghikl}
\item[\em Purpose:]
Call of the main (hard) instanton process generator {\bf QIHGEN}.\\
\item[\em Remarks:]
Replaces a dummy stub in HERWIG~\cite{herwig} that was originally used
as event generation interface for the Monte Carlo generator 
HERBVI~\cite{herbvi} for baryon number violating interactions. 
Used in the QCDINS package to select QCD-instanton induced processes
via MOD\,(IPROC/100,100) $>$ 75. The process code IPROC (= 17600) has to be set
in the user's steering program (c.\,f. Appendix A).  
Furthermore, the QCDINS program header is printed.
\end{nlist}

\noindent
SUBROUTINE {\bf HWBJCO} 
\begin{nlist}{abcdefghikl}
\item[{\em Purpose:}] Modification of HERWIG~\cite{herwig}
                     routine to account for instanton-induced
                      scattering.\\
\item[{\em Remarks:}] The modifications are~\cite{tc}: 
           The logical flag (DISPRO) for keeping the lepton
           momenta fixed in HERWIG 5.9 is modified to include also
           instanton-induced DIS, IPRO = 76. Furthermore,  
           a bug concerning energy-momentum conservation in the original
           routine from HERWIG 5.9 has been fixed.
\end{nlist}

\noindent
SUBROUTINE {\bf HWEGAM}\,(IHEP,\,ZMI,\,ZMA,\,WWA) 
\begin{nlist}{abcdefghikl}
\item[{\em Purpose:}] Modification to avoid standard generation of the 
(virtual) photon at this stage.\\
\item[\em Arguments:] c.\,f. Ref.~\cite{herwig}
\item[\em Remarks:]
This is a modified routine from HERWIG 5.9 (c.\,f. Ref.~\cite{herwig}). 
Usually,  {\bf HWEGAM} generates an incoming photon from the incoming $e^\pm$.
Within QCDINS, however, the photon is generated at a later stage in
the subroutine {\bf QIKGAM}. Thus, the HERWIG routine {\bf HWEGAM} has  
been modified to immediately return for instanton-induced processes (IPRO=76).
\end{nlist}

\noindent
FUNCTION {\bf LAMBERTW}\,(X) 
\begin{nlist}{abcdefghikl}
\item[{\em Purpose:}]
Calculation of the principal branch of the Lambert W-function $W(x)$
for $x \ge 0$.\\ 
\item[{\em Arguments:}] X: $x\ge 0$; argument of the Lambert W-function.
\item[\em Procedure:]
\end{nlist}
$W(x)$ is the (real) solution of
$W(x)\exp(W(x))=x$, analytic at $x=0$.
The following simple, but accurate approximation is used and returned
by {\bf LAMBERTW}:
\begin{equation}
W(x)\approx \left\{
\begin{array}{ll}
0.665\cdot \left(1+0.0195\cdot\ln (x+1)\right)
\ln(x+1)&+\,0.04;\\ 
&{\rm\ for\ }0\leq x\leq 500;\\[0.2cm] 
\ln(x-4)-(1-\frac{1}{\ln(x)})\cdot\ln(\ln(x));& {\rm\ for\ }x > 500\,. 
\end{array} \right.
\end{equation}

\noindent
FUNCTION {\bf LAMNF}\,(NF,\,LAMBDA5) 
\begin{nlist}{abcdefghikl}
\item[\em Purpose:]
Calculation of $\Lambda_{\overline{\rm MS}}^{(n_f<5)}$ from
$\Lambda_{\overline{\rm MS}}^{(5)}$ to order $\alpha_{\overline{\rm MS}}^3$.\\ 
\item[\em Arguments:]\hfill

\begin{nlist}{abcdefghijklm}
        \item[NF:] number of (light) flavours, $n_f$.
        \item[LAMBDA5:] input value $\Lambda_{\overline{\rm MS}}^{(5)}$.
\end{nlist}
\item[\em Procedure:]
The flavour reduction of $\Lambda_{\overline{\rm MS}}^{(5)}$ to the
desired number of light flavours is performed by using Eq.~(9.7) of 
Ref.~\cite{pdg}.  
\end{nlist}

\noindent
SUBROUTINE {\bf LUHEPC}\,(MCONV) 
\begin{nlist}{abcdefghikl}
\item[\em Purpose:]
Conversion of the JETSET~\cite{jetset} event 
record contents back to the HEPEVT common block.\\
\item[{\em Arguments:}] MCONV = 1 
\item[{\em Remarks:}] The present routine is a modified~\cite{ggrind}
version of the JETSET routine {\bf LUHEPC}. 
\end{nlist}

\noindent
FUNCTION {\bf OMEGA}\,(XI4) 
\begin{nlist}{abcdefghikl}
\item[\em Purpose:]
Calculation of the fermionic overlap, as function of the conformally invariant
$\iai$-distance.\\ 
\item[\em Arguments:] XI4: conformally invariant $\iai$-distance $\xi$.
\item[\em Procedure:]
\end{nlist}
The following simple, but accurate approximation
for the fermionic overlap~\cite{rs-pl} $\omega$ is used and returned
by {\bf OMEGA}:
\begin{equation}
\omega (\xi )\approx \frac{4}{(\xi +1/2)^{3/2}}.
\end{equation}

\vspace{8ex}
\noindent
FUNCTION {\bf Q2SIG}\,(XPRIME,\,QLAM,\,KAPPA,\,LOOPFL,\,NF)  
\begin{nlist}{abcdefghikl}
\item[{\em Purpose:}]
Calculation of the total cross section
\begin{equation}
\label{cs}
Q^{\prime 2}\ \sigma^{(I)}_{q^\prime g} \hspace{4ex} [{\rm nb}\ {\rm GeV}^2]
\end{equation}
for the instanton-induced subprocess, depending on 
$x^\prime$, $Q^{\prime}/\Lambda^{(n_f)}_{\overline{\rm MS}}$,
$\mu_r/Q^\prime$,  loop-order and $n_f$. Here, $\mu_r$ and $n_f$
denote the renormalization scale and the number of light flavours,
respectively.\\ 
\item[\em Arguments:]\hfill
\begin{nlist}{abcdefghi}
\item[XPRIME:] $x^\prime$
\item[QLAM:] $Q^{\prime}/\Lambda^{(n_f)}_{\overline{\rm MS}}$
\item[KAPPA:] $\mu_r/Q^\prime$
\item[LOOPFL:] 
\begin{nlist}{abcd}
\item[= 1:] 1-loop renormalization group (RG) invariance~\cite{rs-pl} along 
            with 1-loop form of $\alpha_s$.
\item[= 2:] 2-loop RG invariance~\cite{rs-pl} along with
            2-loop form of $\alpha_{\overline{\rm MS}}$.
\item[= 3:] (default) 2-loop RG invariance along with
            3-loop form of $\alpha_{\overline{\rm MS}}$.
\end{nlist}
\item[NF:] $n_f$; number of light flavours.
\end{nlist}
\item[\em Procedure:]
\end{nlist}
The function {\bf Q2SIG} calculates and returns the cross
section~(\ref{cs}) as derived in  
Ref.~\cite{rs-pl}, 
\begin{eqnarray}
\label{qgcross-vxi}
&&Q^{\prime\,2}\,\sigma_{q^\prime g}^{(I )} =
d^2_{\overline{\rm MS}}\frac{\sqrt{12}}{2^{16}}\pi^{15/2}
\frac{((\xi_\ast +2) v^{\ast\,2}+4 \ts (\ts-2 v_\ast))}
{(v_\ast -\ts)^{9/2}\sqrt{(\xi_\ast +2)v_\ast -4\ts}}
\left( \frac{(\xi_\ast -2)}{\xi_\ast} \frac{\Delta_1\beta_0}{\dts}\right)^{7/2}
\nonumber\\[1.6ex] &&\times
\frac{\omega (\xi_\ast )^{2n_f-1}(\xi_\ast -2)^3 v^{\ast\,5}}
{\sqrt{\frac{1}{2}(\ts -v_\ast -2 \dts )^2 + \ts (\ts -v_\ast ) \ddts}} 
\\[1.6ex] &&\times
\left( \frac{4\pi}
{\alpha_{\overline{\rm MS}}\left(\mu _r \right)}\right)^{19/2}
\exp \left[ 
-\frac{4\pi}
{\alpha_{\overline{\rm MS}}\left(\mu _r \right)} S^{(\iai )}\left(\xi_\ast 
\right) - 2
\left(1-\ln\left(\frac{v_\ast\mu_r}{Q^\prime}\right)\right)
\,\ts\right] \, .
\nonumber
\end{eqnarray}
It is 
expressed entirely in terms of the saddle point values for the $\iai$ 
collective coordinates, $\xi_\ast$ (conformally invariant distance) and
$v_\ast\equiv Q^\prime\rho_\ast$ (scaled size). For given $\xpr$,
$Q^{\prime}/\Lambda^{(n_f)}_{\overline{\rm MS}}$ and (scaled)
renormalization scale $\mu_r/Q^\prime$, these are in turn unique solutions
of the saddle  point equations~\cite{rs-pl}
\begin{eqnarray}
\label{eq2}
\xi_\ast -2 &=& 4\,\frac{\xpr}{1-\xpr}\left
(1-\frac{\ts(\xi_\ast)}{v_\ast}\right)^2,\\[1ex]
\label{eq1}
v_\ast  &=& 2\, D(\ts(\xi_\ast))\, 
W\left( \frac{Q^\prime}{\mu_r}
\frac{\exp\left\{\frac{1}{2}
\left[\frac{4\pi}
{\alpha_{\overline{\rm MS}}\left(\mu _r \right)}\frac{1}{\Delta_1\beta_0} +
\frac{\ts(\xi_\ast)}{D(\ts(\xi_\ast))}
\right]
\right\}}
{2\,D(\ts(\xi_\ast))}
\right),
\end{eqnarray}
with
\begin{equation}
\label{Delta1}
\Delta_1 = \left\{ 
\begin{array}{clcl}
1 & {\rm \ for\ LOOPFL}&=& 1;\\
1+\frac{\beta_1}{\beta_0}
\frac{\alpha_{\overline{\rm MS}}(\mu_r)}{4\pi} 
& {\rm \ for\ LOOPFL}&=& 2,3; 
\end{array}
\right.
\end{equation}
and 
\begin{equation}
\label{Delta2}
\Delta_2 = \left\{
\begin{array}{clcl}
0 & {\rm \ for\ LOOPFL}&=& 1;\\
12\,\beta_0
\frac{\alpha_{\overline{\rm MS}}(\mu_r)}{4\pi} 
& {\rm \ for\ LOOPFL}&=& 2,3. 
\end{array}
\right.
\end{equation}
The $\iai$-action $S^{(\iai )}(\xi )$ as well its $\xi$-derivatives, 
entering the cross section~(\ref{qgcross-vxi}) and Eqs.~(\ref{eq2}),
(\ref{eq1}) through 
\begin{equation}
\label{dadada}
\ts (\xi_\ast )\equiv \Delta_1 \beta_0 S^{(\iai )}(\xi_\ast )-\Delta_2\,;
\hspace{6ex}
D(f(\xi_\ast ))\equiv 
\frac{d}{d\ln (\xi_\ast -2)}f(\xi_\ast )\, ,
\end{equation}
are calculated in the subroutine {\bf ACTION}.  
The fermionic overlap $\omega (\xi)$ is calculated and returned by the 
function {\bf OMEGA}.
In Eq.~(\ref{eq1}), $W$ denotes the principal branch of the Lambert
$W$-function, i.e. the (real) solution of $W(x)\exp(W(x))=x$, analytic
at $x=0$. The latter is calculated and returned by the function {\bf
LAMBERTW}.  

The first step in the solution of the saddle-point
equations~(\ref{eq2}), (\ref{eq1}) consists in eliminating $v_\ast$ in 
Eq.~(\ref{eq2}) by inserting Eq.~(\ref{eq1}).
Next, for given $\xpr$,
$Q^\prime/\Lambda^{(n_f)}_{\overline{\rm MS}}$ and  $\mu_r/Q^\prime$,
the  resulting implicit equation is solved numerically for
$\xi_\ast$. This is done by the function {\bf XI} which
provides
$\xi_\ast=\xi_\ast(\xpr,\,Q^\prime/\Lambda^{(n_f)}_{\overline{\rm MS}},\,
\mu_r/Q^\prime)$ on return. The latter is then 
inserted into Eq.~(\ref{eq1}) providing  $v_\ast=v_\ast 
(\xpr,\,Q^\prime/\Lambda^{(n_f)}_{\overline{\rm MS}},\, \mu_r/Q^\prime)$. 

The values of $4\pi/\alpha_{\overline{\rm MS}}$ are calculated
and returned by the function {\bf XQS}. 

\noindent
SUBROUTINE {\bf QCDGEN} 
\begin{nlist}{abcdefghikl}
\item[\em Purpose:] Interface for the generation of  one
instanton-induced event, including calls to event initialization,
hadronization  and event termination routines.\\ 
\item[\em Procedure:]
\end{nlist}
\begin{itemize}
\item An instanton-induced event is initialized by a call to the 
      HERWIG~\cite{herwig} subroutine {\bf HWUINE}. 
\item The partonic instanton subprocess is generated by a call to
      the HERWIG subroutine {\bf HWEPRO}.   
\item If the hadronization flag  QICONT(21) is
      set .TRUE.  (default), the event is fully hadronized. Else,
      the event is finalized ({\bf HWUFNE}) immediately after the call 
      of {\bf HWEPRO}.  
      Depending on the control flag for the hadronization model, 
      QICONT(18)=.TRUE./.FALSE., the hadronization is either performed
      by appropriate HERWIG~\cite{herwig} or 
      (modified~\cite{HJung}) JETSET~\cite{jetset} routines, respectively (see 
      Table~\ref{hadroutines}).  
\item Furthermore, in this routine energy-momentum
      conservation in the generated event is checked.
\end{itemize}
\begin{table}
\begin{center}
\begin{tabular}{|c||l|p{7.5cm}|}\hline
QICONT(18)& \multicolumn{2}{c|}{Hadronization scheme}\\\hline\hline
& \multicolumn{2}{c|}{HERWIG Interface} \\ \cline{2-3}\cline{2-3}
&{\bf HWUINE} & Initialize event\\
&{\bf HWEPRO} & Generate hard subprocess\\
&{\bf HWBGEN} & Generate parton cascades\\
&{\bf HWDHQK} & Do heavy quark decays\\
&{\bf HWCFOR} & Do cluster hadronization\\
\rb{.TRUE.}&{\bf HWCDEC} & Do cluster decay\\
&{\bf HWDHAD} & Do unstable particle decays\\
&{\bf HWDHVY} & Do heavy flavour decays\\
&{\bf HWMEVT} & Add soft underlying event if needed\\
&{\bf HWUFNE} & Finalize event\\
\hline\hline
& \multicolumn{2}{c|}{JETSET  Interface} \\ \cline{2-3}\cline{2-3}
&{\bf HWUINE} & Initialize event\\
&{\bf HWEPRO} & Generate hard subprocess\\
&{\bf HWBGEN} & Generate parton cascades\\
&{\bf HWUFNE} & Finalize event\\
&{\bf HERLUND} & Convert HERWIG event record in HEPEVT common block 
        to the respective JETSET common block\\
.FALSE.&{\bf LUGIVE} & Assign the hard subprocess (HARD) and instanton (INST)
       character strings in JETSET common block\\
&{\bf LUEXEC} & Simulate the whole fragmentation and decay chain\\
&{\bf LULIST}(1) & List the (first) event record\\
&{\bf LUHEPC}(1) & Convert JETSET event record contents back to  
          the HEPEVT common block\\
\hline
\end{tabular}
\end{center}
\vspace{3ex}
\caption[dum]{\label{hadroutines} 
Depending on the flag QICONT(18) = .TRUE./.FALSE., HERWIG~\cite{herwig} and
JETSET~\cite{jetset} hadronization may be selected. The respective
calls to HERWIG and (modified) JETSET routines used in {\bf QCDGEN}
are displayed.}   
\end{table}

\noindent
SUBROUTINE {\bf QCLOOP} 
\begin{nlist}{abcdefghikl}
\item[{\em Purpose:}] Loop over calls to the instanton event
generator and interface of the instanton-induced hard
subprocess to HERWIG~\cite{herwig} initialization and output
routines.\\   
\item[\em Procedure:]
\end{nlist}
\begin{itemize}
\item  The instanton-induced hard subprocess is initialized
by calling the HERWIG subroutine {\bf HWEINI}. 
\item The generator {\bf QCDGEN} for an instanton-induced event 
      and, thereafter, the user's routine {\bf HWANAL} for analyzing data from
      the event, are called in a loop MAXEV times, with MAXEV being the desired
      number of events. MAXEV has to be set in the user's steering program
      (c.\,f. Appendix A).
\item The instanton-induced hard subprocess is terminated by
      calling the HERWIG subroutine {\bf HWEFIN} that produces the
      final output.  
\end{itemize}

\noindent
SUBROUTINE {\bf QICALC} 
\begin{nlist}{abcdefghikl}
\item[{\em Purpose:}] Calculates derived parameters and checks the
                    chosen kinematical boundaries. The latter are
                    adjusted in case of inconsistencies.\\ 
\item[\em Non-local variables  (re)set:]\hfill
\begin{nlist}{abcdefghiklmnop}
\item[QILPAR(I):] = LOG(QIUPAR(I)), I = 1\,..\,18, if
                  QIUPAR(I) $> 1.0\cdot 
                  10^{-20}$; otherwise QILPAR(I) $= -1.0\cdot 10^{25}$; 
                  these parameter settings are partially reset after exit 
                  (see below).
\item[QIDPAR(1):] = QIUPAR(12)**2; square of minimum allowed instanton CM
                  energy $W_{I\,{\rm min}}^2$.
\item[QIWARF(1):] = .FALSE. (default); if $Q^{\prime 2}_{\rm max}$ violates 
                  the kinematical constraint 
                  \begin{equation}
                   \label{warn1}
                   Q^{\prime 2}_{\rm max}\leq 
                   z_{\rm max} S-W_{I\,{\rm min}}^2 ,
                  \end{equation}
                  QIWARF(1) is reset to .TRUE..
\item[QINWAR(1):] = 0 (default); counts the number of violations of 
                   Eq.~(\ref{warn1}). 
\item[QILPAR(8):] $=\ln Q^{\prime 2}_{\rm max}$ (default); reset to the 
                  logarithm of the right hand side of Eq.~(\ref{warn1}) if 
                  this kinematical constraint is violated.
\item[QIUPAR(8):] $=Q^{\prime 2}_{\rm max}$; reset to the right hand side of
                  Eq.~(\ref{warn1}) if this kinematical constraint is violated.
\item[QIWARF(2):] = .FALSE. (default); if $Q^{\prime 2}_{\rm min}$ violates 
                  the kinematical constraint 
                  \begin{equation}
                   \label{warn2}
                   Q^{\prime 2}_{\rm min}\geq 
                   \frac{x_{\rm Bj\, min}}{z_{\rm max}-x_{\rm Bj\, min}}
                  W_{I\,{\rm min}}^2 ,
                  \end{equation}
                  QIWARF(2) is reset to .TRUE..
\item[QINWAR(2):] = 0 (default); counts the number of violations of 
                   Eq.~(\ref{warn2}). 
\item[QILPAR(9):] $=\ln Q^{\prime 2}_{\rm min}$ (default); reset to the 
                  logarithm of the right hand side of Eq.~(\ref{warn2}) if 
                  this kinematical constraint is violated.
\item[QIUPAR(9):] $=Q^{\prime 2}_{\rm min}$; reset to the right hand side of
                  Eq.~(\ref{warn2}) if this kinematical constraint is violated.
\item[QIWARF(3):] = .FALSE. (default); if $x^\prime_{\rm max}$ violates 
                  the kinematical constraint 
                  \begin{equation}
                   \label{warn3}
                   x^\prime_{\rm max}\leq 
                  1-\frac{W_{I\,{\rm min}}^2}{z_{\rm max} S},
                  \end{equation}
                  QIWARF(3) is reset to .TRUE..
\item[QINWAR(3):] = 0 (default); counts the number of violations of 
                   Eq.~(\ref{warn3}). 
\item[QILPAR(6):] $=\ln x^{\prime}_{\rm max}$ (default); reset to the 
                  logarithm of the right hand side of Eq.~(\ref{warn3}) if 
                  this kinematical constraint is violated.
\item[QIUPAR(6):] $=x^{\prime}_{\rm max}$; reset to the right hand side of
                  Eq.~(\ref{warn3}) if this kinematical constraint is violated.
\item[QIWARF(4):] = .FALSE. (default); if $x^{\prime}_{\rm min}$ violates 
                  the kinematical constraint 
                  \begin{equation}
                   \label{warn4}
                   x^{\prime}_{\rm min}\geq 
                  \frac{x_{\rm Bj\,min}}{z_{\rm max}},
                  \end{equation}
                  QIWARF(4) is reset to .TRUE..
\item[QINWAR(4):] = 0 (default); counts the number of violations of 
                   Eq.~(\ref{warn4}). 
\item[QILPAR(7):] $=\ln x^{\prime}_{\rm min}$ (default); reset to the 
                  logarithm of the right hand side of Eq.~(\ref{warn4}) if 
                  this kinematical constraint is violated.
\item[QIUPAR(7):] $=x^{\prime}_{\rm min}$; reset to the right hand side of
                  Eq.~(\ref{warn4}) if this kinematical constraint is violated.
\item[QIWARF(11):] = .FALSE. (default); reset to .TRUE. if current
                  quark mass requirement (c.\,f. QICONT(6)) does not
                  fit to the selected order of variable generation 
                  (c.\,f. QICONT(5)).
\item[QINWAR(11):] = 0 (default); counts the number of recurring QIWARF(11) =
                  .TRUE. settings.
\item[QICONT(6):] = .TRUE. (default); reset to .FALSE. if the flag 
                  QIWARF(11) = .TRUE..
\item[QIQMASS(1,\,I):] $\sum_{i=1}^{I} m_{q_i}$; cumulative sum of
HERWIG~\cite{herwig} quark masses with $1\le {\rm I}\le 6$ referring
to the HERWIG identity code.  
\item[QIQMASS(2,\,I):] $m_{q_i}$; HERWIG quark mass with $1\le {\rm
I}\le 6$ referring to the HERWIG identity code.
\item[QIRCAL:] = .TRUE. (default); reset to .FALSE. on exit.\\
\end{nlist}
\item[\em Remarks:] This routine is only performed if the flag QIRCAL
     is set to .TRUE. ({\bf QIINIT}, {\bf QISETD}, {\bf QISETF}, 
     {\bf QISETI}). 
\end{nlist}

\noindent
SUBROUTINE {\bf QICCON} 
\begin{nlist}{abcdefghikl}
\item[{\em Purpose:}]
Assignment of the colour and flavour connections for the partons.\\ 
\item[\em Non-local variables  (re)set:]\hfill
\begin{nlist}{abcdefghiklmnopqrst}
\item[QIPLIS\,(JLP,\,ILP):] IHEP pointer referring to the
             outgoing parton JLP in the string ILP. 
\item[JMOHEP:] array of HERWIG~\cite{herwig} ``mother'' pointers. 
\item[JDAHEP:] array of HERWIG ``daughter'' pointers.\\
\end{nlist}
\item[\em Remarks:]
For an explicit example, see  the description of routine {\bf QISTID}.
\item[\em Procedure:]
\end{nlist}
\begin{itemize}
\item The IHEP pointer of the current quark $k$  
      (IHEP=10), as assigned in {\bf QIHGEN}, is added to the array 
      QIPLIS of IHEP pointers at the position (JLP,ILP) of the 
      virtual quark $q^\prime$.
\item The IHEP pointer of the incoming gluon (IHEP=6), as assigned 
      in {\bf QIHGEN}, is added to the array QIPLIS of IHEP pointers
      at the position (JLP,ILP) of the incoming gluon.
\item The colour flow is constructed by connecting the colour lines of 
      neighbouring partons within each of the strings constructed in the 
      subroutines {\bf QIGLST} and {\bf QIGPAR}.
\item Finally, the flavour flow is constructed by connecting the flavour lines
      of the quark at the beginning of a string with the flavour line
      of the anti-quark at the end of a string. 
\end{itemize}

\noindent
SUBROUTINE {\bf QIGETD}\,(NUM,\,VALUE,\,OK)
\begin{nlist}{abcdefghikl}
\item[\em Purpose:]
Get value of double precision parameter QIUPAR(NUM) for given NUM.\\  
\item[\em Arguments:]\hfill
\begin{nlist}{abcdefghi}
\item[NUM:] integer, pointing to the parameter QIUPAR(NUM) to be read.     
\item[VALUE:] value of double precision parameter QIUPAR(NUM).      
\item[OK:] logical return flag.    
\end{nlist}
\item[\em Non-local variables (re)set:]\hfill
\begin{nlist}{abcdefghiklmno}
\item[QINWAR(10):] = QINWAR(10)+1; counts attempts to read non-existing
                    parameter QIUPAR(.).
\item[QIWARF(10):] = .TRUE. if parameter index NUM out of range.
\end{nlist}
\item[\em Remarks:]
Service routine for reading out the double precision input
parameters QIUPAR(NUM), NUM=1,2,\ldots,18. On return, the logical flag OK is
.TRUE. if NUM is within its allowed range.  
\end{nlist}

\noindent
SUBROUTINE {\bf QIGETF}\,(NUM,\,FLAG,\,OK)
\begin{nlist}{abcdefghikl}
\item[\em Purpose:]
Get value of logical flag QICONT(NUM) for given NUM.\\
\item[\em Arguments:]\hfill
\begin{nlist}{abcdefghi}
\item[NUM:] integer, pointing to the parameter QICONT(NUM) to be
            read.                   
\item[FLAG:] boolean value of QICONT(NUM).       
\item[OK:]  logical return flag. 
\end{nlist}
\item[\em Non-local variables (re)set:]\hfill
\begin{nlist}{abcdefghiklmno}
\item[QINWAR(6):] = QINWAR(6)+1; counts attempts to read non-existing
                    parameter QICONT(.).
\item[QIWARF(6):] = .TRUE. if parameter index NUM out of range.
\item[QIRCAL:] = .TRUE. if QIWARF(6) = .FALSE. (default).
\end{nlist}
\item[\em Remarks:]
Service routine for reading out the boolean input parameters
QICONT(NUM), NUM=1,2,\ldots,21.
On return, the logical flag OK is .TRUE. if NUM is within its
allowed range. 
\end{nlist}

\noindent
SUBROUTINE {\bf QIGETI}\,(NUM,\,VALUE,\,OK)
\begin{nlist}{abcdefghikl}
\item[\em Purpose:]
Get value of integer parameter QIPARI(NUM) for given NUM.\\ 
\item[\em Arguments:]\hfill
\begin{nlist}{abcdefghi}
\item[NUM:] integer, pointing to the parameter QIPARI(NUM) to be
read. 
\item[VALUE:] value of QIPARI(NUM).           
\item[OK:] logical return flag. 
\end{nlist}
\item[\em Non-local variables (re)set:]\hfill
\begin{nlist}{abcdefghiklmno}
\item[QINWAR(8):] = QINWAR(8)+1; counts attempts to read non-existing
                    parameter QIPARI(.).
\item[QIWARF(8):] = .TRUE. if parameter index NUM out of range.
\item[QIRCAL:] = .TRUE. if QIWARF(8) = .FALSE. (default).
\end{nlist}
\item[\em Remarks:]
Service routine for reading out the integer input
parameters QIPARI(NUM), NUM=1,2,\ldots,10. 
On return, the logical flag OK is .TRUE. if NUM is within 
its allowed range. 
\end{nlist}

\noindent
SUBROUTINE {\bf QIGLST} 
\begin{nlist}{abcdefghikl}
\item[{\em Purpose:}]
Generation of $n_f$ $[\,q \ldots \overline{q}\,]$ - ``strings'' of
partons, with a total number $n_g+1$ of gluons inserted in between, in
accord with  the requirement of  ``flavour democracy'' among the
$n_f$ $q\overline{q}$-pairs  participating in the instanton-induced
subprocess.\\ 
\item[\em Non-local variables  (re)set:]\hfill
\begin{nlist}{abcdefghiklmnopqrst}
\item[QIPTYP\,(JLP,\,ILP):]  particle data group~\cite{pdg} identity code  
(IDPDG) of the parton JLP in the string ILP.   
\item[QINLIS\,(ILP):] number of partons in the string ILP.
\item[QIPINF\,(JLP,\,ILP):] 
\begin{nlist}{abcdefghikl}
\item[= .FALSE.;] logical flag indicating whether parton JLP in the
string ILP is incoming.
\end{nlist}
\end{nlist}
\item[\em Procedure:]
\end{nlist}
Each of the $n_f$ generated strings begins with a quark, followed by a random
number $\le n_g+1$ of gluons, and ends with an anti-quark of randomly
chosen flavour.  Due to the required ``flavour democracy'',
$q\overline{q}$-pairs of  all $n_f$ flavours must occur precisely once, 
e.\,g. for $n_f=3$ and $n_g=3$, 
\begin{equation}
[\,d\; g\;\overline{u}\,]\hspace{5ex} [\,u\;g\;g\;\overline{s}\,]
\hspace{5ex} 
[\,s\;g\;\overline{d}\,]\, .
\end{equation} 
In practice, the strings are generated and administrated as 
follows:
\begin{itemize}
\item First,  a 2-dimensional integer array QIPTYP is defined, whose entries
QIPTYP\,(JLP,\,ILP) will contain the  particle data group identity code  
      (IDPDG) of the parton JLP in the ILP-th $[\,q \ldots
      \overline{q}\,]$ - string. All entries are initially set to zero.
\item The quarks, labelled by their IDPDG-codes $1\,..\,n_f$, are then
      inserted consecutively at the top of the first $n_f$ columns 
      (``strings''), e.\,g. for $n_f=3$, 
{\renewcommand{\arraystretch}{1.2}
      $$ \begin{array}{cr|rrrrrr}
                 &    & \multicolumn{6}{|c}{{\rm ILP}}\\ 
                 &    & 1 & 2 & 3 & 4 &\ .\ & 30\\ \hline
                 & 1\  & 1 & 2 & 3 & 0 &\ .\ & 0 \\
                 & 2\  & 0 & 0 & 0 & 0 &\ .\ & 0 \\
 {\rm JLP}       & .\  & . & . & . & . &\ .\ & . \\
                 & 30\ & 0 & 0 & 0 & 0 & \ .\ & 0 \\ \hline
       {\rm QINLIS(ILP)} &  & 1 & 1 & 1 & 0 & \ .\ & 0 \\ 
      \end{array}$$
      }
\item Next, the anti-quarks with IDPDG-codes   
      $-1\,..\,-n_f$, are inserted randomly in the second row of the
      first $n_f$ columns, e.g.
{\renewcommand{\arraystretch}{1.2}
      $$\begin{array}{cr|rrrrrr}
                 &    & \multicolumn{6}{|c}{{\rm ILP}}\\ 
                 &    & 1 & 2 & 3 &\ 4 &\ .\ & 30\\ \hline
                 & 1\  & 1 & 2 & 3 &\ 0 &\ .\ & 0 \\
                 & 2\  &-2 &-3 &-1 &\ 0 &\ .\ & 0 \\
  {\rm JLP}      & 3\  & 0 & 0 & 0 &\ 0 &\ .\ & . \\
                 & .\  & . & . & . &\ . &\ .\ & . \\
                 & 30\ & 0 & 0 & 0 &\ 0 &\ .\ & 0 \\ \hline
       {\rm QINLIS(ILP)} &  & 2 & 2 & 2 &\ 0 &\ .\ & 0 \\ 
      \end{array}$$
       }
\item Finally, the $n_g+1$ gluons of the instanton 
      subprocess with IDPDG-code 21 are 
      distributed randomly over the first $n_f$ columns in between the 
      quarks and anti-quarks: For each of the gluons, one of the first $n_f$
      columns is picked at random,   
      the anti-quark in this column is moved one row further down and
      the gluon is inserted at the anti-quark's  previous position. 
      A possible outcome of this procedure for the example above
      with $n_g=3$ is
{\renewcommand{\arraystretch}{1.2}
      $$\begin{array}{cr|rrrrrr}
                 &    & \multicolumn{6}{|c}{{\rm ILP}}\\ 
                 &    & 1 & 2 & 3 &\ 4 & \ .\ & 30\\ \hline
                 & 1\  & 1 & 2 & 3 &\ 0 & \ .\ & 0 \\
                 & 2\  &21 & 21 & 21 &\ 0 & \ .\ & 0 \\
                 & 3\  &-2 & 21 & -1 &\ 0 & \ .\ & . \\
      {\rm JLP}  & 4\  &0  & -3 & 0  &\ 0 & \ .\ & . \\
                 & 5\  & 0 & 0  & 0  &\ 0 & \ .\ & . \\
                 & .  & . & .  & .  &\ . & \ .\ & . \\
                 & 30 & 0 & 0  & 0  &\ 0 & \ .\ & 0 \\ \hline
       {\rm QINLIS(ILP)} &  & 3 & 4 & 3 &\ 0 & \ .\ & 0 \\ 
      \end{array}$$
       }
\end{itemize}

\noindent
SUBROUTINE {\bf QIGMUL}\,(WGTFCT) 
\begin{nlist}{abcdefghikl}
\item[\em Purpose:]
Generation of the number of gluons $n_g$ emitted by the instanton 
subprocess in accord with kinematical constraints.\\
\item[\em Arguments:]\hfill
 
\begin{nlist}{abcdefghiklmno}
\item[WGTFCT:] 
\begin{nlist}{abcde}
\item[$= 1.0$,] returned, if the event is to be accepted.
\item[$= 0.0$,] returned, if the final state quark masses are too
high; event will be killed (default).
\end{nlist}
\end{nlist}
\item[\em Non-local variables (re)set:]\hfill
\begin{nlist}{abcdefghiklmno}
\item[QINGLU:] $n_g+1$, number of gluons involved in the instanton subprocess.
\item[QIWARF(14):] = .FALSE. (default); reset to .TRUE. if the
calculation of the number of gluons requires too many iterations.
\item[QINWAR(14):] = 0 (default); counts the number of recurring QIWARF(14) =
                  .TRUE. settings. 
\end{nlist}
\item[\em Procedure:]
\end{nlist}
\begin{itemize}
\item The minimal CM energy available for gluon generation is
      calculated and WGTFCT is set accordingly.
\item If the emission of at least one gluon is energetically allowed,
      the average gluon multiplicity $\langle n_g\rangle^{(I)}$ is 
      calculated and returned by the function {\bf GMULT}. 
\item The number $n_g$ of emitted gluons is then generated
     according to a Poisson distribution~\cite{dis97} with mean 
     $\langle n_g\rangle^{(I)}$,
      \begin{equation}
      P(n_g) = \frac{\langle n_g\rangle^{(I)\ n_g}}{n_g!}\, 
              {\rm e}^{-\langle n_g\rangle^{(I)}} .
      \end{equation} 
\item An upper limit on $\langle n_g \rangle^{(I)}$ is enforced and the
      kinematical constraints are checked again. In
      case they are violated, a new iteration cycle is started
      corresponding to a different generated value of $n_g$.    
\end{itemize}

\noindent
SUBROUTINE {\bf QIGPAR} 
\begin{nlist}{abcdefghikl}
\item[\em Purpose:]
Identification of the incoming partons among the  partons
participating  in the instanton-induced subprocess.\\  
\item[\em Non-local variables (re)set:]\hfill
\begin{nlist}{abcdefghiklmnopqrs}
\item[QIPINC(1,1):] row number (JLP) in the 2-dimensional array 
QIPTYP ({\bf QIGLST}), where the anti-parton corresponding
to the incoming virtual (anti-)quark $q^\prime$ is located. 
\item[QIPINC(1,2):] column number (ILP) in the 2-dimensional array 
QIPTYP ({\bf QIGLST}), where the anti-parton corresponding
to the incoming virtual (anti-)quark $q^\prime$ is located.
\item[QIPINC(2,1):] row number (JLP) of the incoming on-shell gluon $g$
in the 2-dimensional array QIPTYP ({\bf QIGLST}).
\item[QIPINC(2,2):] column number (ILP) of the incoming on-shell gluon $g$
in the 2-dimensional array QIPTYP ({\bf QIGLST}).
\item[QIPINF(JLP,\,ILP):]  =\,.FALSE. (default); set to .TRUE. only if
(JLP, ILP) = (QIPINC(I,1),QIPINC(I,2)), with I=1,2, corresponding to
the incoming partons.\\   
\end{nlist}
\item[\em Procedure:]
\end{nlist}
One of the $n_g+1$ gluons in the generated $n_f$  $[\,q \ldots
\overline{q}\,]$-strings ({\bf QIGPAR}) is randomly chosen and marked
as incoming.  Furthermore, the anti-parton corresponding to the
virtual (anti-)quark $q^\prime$ is identified and marked as incoming.

\noindent
SUBROUTINE {\bf QIHGEN} 
\begin{nlist}{abcdefghikl}
\item[{\em Purpose:}] Main (hard) instanton process generator.\\ 

\item[\em Non-local variables (re)set:]\hfill

\begin{nlist}{abcdefghiklmno}
\item[QISGAM:]  $eP$ CM energy squared $S$.
\item[IDN(1):]  HERWIG identity code (IDHW) of incoming $e$.
\item[IDN(2):]  HERWIG identity code (IDHW) of incoming $g$.
\item[EMSCA:]   factorization scale $\mu_f=$ QIUPAR(18)
\item[QIZPAR:]  generated $z$.
\item[QIXONE:]  $x^\prime\,z$ 
\item[QIYONE:]  $Q^{\prime 2}/(S x^\prime z )$
\item[QIXBJG:]  generated $x_{\rm Bj}$.
\item[QIYBJG:]  generated $y_{\rm Bj}$.
\item[QIQ2GA:]  Bjorken variable $Q^2$ calculated via 
                $Q^2=S x_{\rm Bj} y_{\rm Bj}$.
\item[QIELEN:]  energy of incoming $e$, $e_t$.
\item[QIPREN:]  energy of incoming $P$, $P_t$.
\item[EVWGT:]   weight $W_{eP}^{(I)}$~(\ref{EVWGT}) corresponding to the 
instanton-induced cross section~(\ref{evwgt}).
\item[QIMEWT:] internal storage of weight corresponding to Eq.~(\ref{evwgt}) 
for analysis 
\item[QIPINC(1,3):] particle data group identity code (IDPDG) of virtual quark 
$q^\prime$.
\item[QIPINC(2,3):] particle data group identity code (IDPDG) of incoming 
                gluon $g$. 
\item[QIPINC(1,4):] (=10) IHEP pointer of current quark $k$.
\item[QIPINC(2,4):] (=6) IHEP pointer of incoming $g$.
\item[QISHEP:] CM energy squared of instanton subprocess, $W_I^2=(q^\prime +
g)^2=s^\prime$.
\item[JMOHEP:] array of HERWIG~\cite{herwig} ``mother'' pointers. 
\item[JDAHEP:] array of HERWIG ``daughter'' pointers.\\
\end{nlist} 
\item[\em Procedure:]
\end{nlist}
For the generation of an event, this routine
is called twice: in the first call, a Monte Carlo weight
associated with the instanton-induced total cross section is
calculated. In the second call, the generation of the event is completed.

In the first call (GENEV = .FALSE.), the following steps are performed
in {\bf QIHGEN}:  
\begin{itemize}
\item Information on the incoming beams is accumulated, 
      initial state radiation from the lepton is optionally accounted
      for ({\bf EXFRAC}),  various derived 
      parameters are calculated and kinematical limits are checked
      ({\bf QICALC}). 
\item The HERWIG~\cite{herwig} identity code (IDHW) of the current quark 
      $k$ and of the virtual quark $q^\prime$
      (c.\,f. Fig.~\ref{kin-var}) are generated ({\bf QIHPAR}).  
\item Next follows the generation of the various Bjorken variables
      $Q^{\prime 2}$, $x^\prime$, $z$, $x_{\rm Bj}$, $y_{\rm Bj}$   
      (c.\,f. Fig.~\ref{kin-var}) of the instanton-induced DIS
      process, with weights corresponding
      to the associated factors appearing in the instanton-induced total  
      cross section~(\ref{evwgt}):\\ 
\begin{itemize}
\item[$\circ$] By a call to {\bf QIHINS}, the Bjorken variables $Q^{\prime
      2}$ and $x^\prime$ of the instanton subprocess are
      generated. Moreover,  
      the Monte Carlo weight $W_{Q^\prime x^\prime\sigma}$ from
      Eq.~(\ref{QIWGT}), corresponding to the integral
      \begin{equation}
      \label{qiwgt}
      \int\limits_{Q^{\prime 2}_{\rm min}}^{Q^{\prime 2}_{\rm max}} 
      dQ^{\prime 2}\,
      \int\limits_{x^\prime_{\rm min}}^{x^\prime_{\rm max}}
      \frac{dx^\prime}{x^\prime}
            \frac{2}{x^\prime}\ 
      \sigma^{(I)}_{q^\prime g}(x^\prime, Q^{\prime 2} ),
      \end{equation}
      is calculated, where $\sigma^{(I)}_{q^\prime g}$ denotes the 
      instanton-induced $q^\prime g$ total cross section~\cite{rs-pl}
      from Eq.~(\ref{qgcross-vxi}). At this stage, also the gluon
      multiplicity depending on $x^\prime$ and $Q^{\prime 2}$ is
      generated ({\bf QIGMUL}).   
\item[$\circ$] The fractional momentum $z$ of the incoming
      gluon is generated and the Monte Carlo weight $W_{z g}$  
      corresponding to the integral   
      \begin{equation}
      \label{zweight} 
      \int\limits_{{\rm max}\left(\frac{Q^{\prime 2}}{Sx^\prime 
      y_{\rm Bj\, max}},\frac{x_{\rm Bj\,min}}{x^\prime}
      \right)}^{z_{\rm max}}
      dz\,f_g (z, \mu_f ),
      \end{equation}
      is calculated,
      where $f_g$ is the gluon distribution in the proton.
      By default, $f_g$ is taken from Ref.~\cite{owens} (set 1.1) and the 
      factorization scale $\mu_f$ is identified with $Q^\prime_{\rm min}$.
      For the default setting of the control flags in {\bf QIINIT},
      the value of 
      $z$ as well as the Monte Carlo weight $W_{zg}$ are generated 
      as follows: Random points $y_z$, with 
      \begin{equation}
      \label{yz}
      y_z = \ln z\hspace{3ex} \Leftrightarrow\hspace{3ex} z = \exp y_z
      \end{equation}
      are generated uniformly in the interval between 
      $\ln z_L$ and $\ln z_U$, where $z_L$ and $z_U$ are the lower
      and upper integration limits in Eq.~(\ref{zweight}),
      respectively. Hence, the Monte Carlo weight associated with
      Eq.~(\ref{zweight}) reads 
      \begin{equation}
      \label{Wzg}
      W_{zg} = (\ln z_U - \ln z_L )\,z\,f_g(z,\mu_f) .
      \end{equation}
\item[$\circ$] The Bjorken variables $x_{\rm Bj}$ and $y_{\rm
      Bj}$ are generated analogously to the $z$ variable
      and the weight $W_{x_{\rm Bj}\,y_{\rm Bj}}$,
      \begin{equation}
      \label{Wxybj}
      W_{x_{\rm Bj}\,y_{\rm Bj}} = 
      \ln\frac{x_{{\rm Bj}\,U}}{x_{{\rm Bj}\,L} }\,
      \ln\frac{y_{{\rm Bj}\,U}}{y_{{\rm Bj}\,L} } \, 
      \theta (S x_{\rm Bj} y_{\rm Bj} - Q^2_{\rm min})\,,
      \end{equation}
      corresponding to the 
      integral 
      \begin{equation}
      \label{xybj}
      \int\limits_{x_{{\rm Bj}\,L}}^{x_{{\rm Bj}\,U}} 
      \frac{dx_{\rm Bj}}{x_{\rm Bj}}
      \int\limits_{y_{{\rm Bj}\,L}}^{y_{{\rm Bj}\,U}}
      \frac{dy_{\rm Bj}}{y_{\rm Bj}}\  
      \theta (S x_{\rm Bj} y_{\rm Bj} - Q^2_{\rm min})\,
      \end{equation}
      is calculated.  
      The integration limits in Eq.~(\ref{xybj}) are
      given in terms of the physical $x_{\rm Bj},y_{\rm Bj}$-cuts, with 
      default values as displayed in Table~\ref{floating},
\begin{eqnarray}
x_{{\rm Bj}\,U} &=& x^\prime z -
      \frac{m_{k}^2}{S} 
      \frac{1}{y_{\rm Bj\,max}-\frac{Q^{\prime 2}}{Sx^\prime z}};\nonumber\\
x_{{\rm Bj}\,L} &=& x_{\rm Bj\,min};\nonumber\\
y_{{\rm Bj}\,U} &=& y_{\rm Bj\,max};\\
y_{{\rm Bj}\,L} &=& {\rm max}\left( 
      \frac{Q^{\prime 2}}{Sx^\prime z}+\frac{m_{k}^2}{S}
      \frac{1}{x^\prime z -x_{\rm Bj}},y_{\rm Bj\,min}\right).\nonumber
\end{eqnarray}
\end{itemize}
\item Finally, given the generated Bjorken variables, the weights
      associated with the remaining factors in the instanton-induced
      cross section~(\ref{evwgt}) are calculated:\\
\begin{itemize}     
\item[$\circ$] The weight arising from the splitting of the photon into 
      a $q\overline{q}$-pair is obtained from the routine {\bf QISPLT},
      \begin{equation}
      \label{splitting}
      W_{q^\prime} = \frac{1}{x^\prime z}\, P_{q^\prime}^{{ (I)}}
      (x^\prime ,Q^{\prime 2} ,z,Q^2=S x_{\rm Bj} y_{\rm Bj}).
      \end{equation}
      The factor $P_{q^\prime}^{{ (I)}}$ from Eq.~(\ref{flux})
      accounts for the flux  
      of virtual (anti-)quarks $q^\prime$ in the instanton background,
      entering the instanton-induced $q^\prime g$-subprocess from the photon 
      side~\cite{rs-pl,mrs2} (c.\,f. Fig.~\ref{kin-var}).  
\item[$\circ$] The familiar weight from the flux of virtual photons is
calculated, 
      \begin{equation}
      \label{photonflux}
      W_\gamma = \frac{1}{S x_{\rm Bj} y_{\rm Bj}} 
      \left(1-y_{\rm Bj} +\frac{y_{\rm Bj}^2}{2}\right) .
      \end{equation}
\item[$\circ$] The left-over weight factor required for a
      properly normalized cross section (in units of nb) is obtained
      from the routine {\bf QIPVWT}, 
      \begin{equation}
      \label{crossweightfac}
      W_n = 2\,\pi\,\alpha^2\, x_{\rm Bj}\, x^\prime \,
      \sum\limits_{q^\prime = d,u,s,\ldots ;\,
      \overline{d},\overline{u},\overline{s},\ldots}  
      e_{q^\prime}^2 .
      \end{equation}
      Here $\alpha$ is the electro-magnetic fine-structure constant and
      $e_{q^\prime}$ denotes the fractional electro-magnetic charge of the
      virtual (anti-)quark $q^\prime$. The sum extends over quarks
      and anti-quarks of the considered $n_f$ light flavours.\\  
\end{itemize}
\item The total event weight (EVWGT) required by the HERWIG package, is
      the product of all the above weights, 
      \begin{equation}
      \label{EVWGT}
      W_{eP}^{(I)} = W_{Q^\prime x^\prime\sigma}\,
      W_{z g}\,W_{x_{\rm Bj}\,y_{\rm Bj}}\,W_{q^\prime}\,
      W_\gamma\,W_n .
      \end{equation}
\end{itemize}            
This ends the first call of {\bf QIHGEN}. 

\noindent
For standard settings of the control flags in the HERWIG~\cite{herwig} 
routine {\bf HWIGIN} (NOWGT=.TRUE.), the (standard) rejection method
is applied producing an unweighted event distribution according to the
instanton-induced, normalized differential cross section (\ref{diffcross}).

\vspace{1ex}
If the event is accepted (GENEV = .TRUE.), {\bf QIHGEN} is called
again. The following steps are then performed:
\begin{itemize}
\item The lepton beam $e$ is stored in the event record. 
\item The 4-momentum $g$ of the incoming gluon is
      calculated from its generated fractional momentum $z$ and the 
      4-momentum of the incoming proton ({\bf
      QIKPAR}). The result is stored in the event record.
\item The 4-momentum $q$ of the virtual photon is
      generated ({\bf QIKGAM}).  
\item The 4-momentum $e^\prime$ of the outgoing lepton 
      is calculated from $e$, $q$ and stored. 
\item The 4-momenta of the virtual quark $q^\prime$ and
      the outgoing current quark $k$  are generated 
      ({\bf QIKGSP}) and  copied into the event record. 
\item The required data for the generation routines of the instanton
      subprocess 
       \begin{equation}
       q^\prime g \stackrel{I}{\Rightarrow} X
       \label{isubproc}
       \end{equation}
       are set up and its (total) momentum vector $q^\prime + g$ is
       calculated.   
\item The colour connections of the hard subprocess are initiated. 
\item The instanton-induced partonic final state $X$ in
      Eq.~(\ref{isubproc}) is generated  ({\bf QISTID}) with all
      colour connections in the $q^\prime g$-CM system and, thereafter,
      transformed into the laboratory frame.   
\end{itemize}

\noindent
SUBROUTINE {\bf QIHINS}
\begin{nlist}{abcdefghikl}
\item[{\em Purpose:}]
Generation of the Bjorken variables $Q^{\prime 2}$ and $x^\prime$, 
associated with the instanton subprocess, and calculation of the Monte Carlo 
weight associated with the integral~(\ref{qiwgt}).
\item[\em Non-local variables (re)set:]\hfill
\begin{nlist}{abcdefghik}
\item[QIL2IN:] $Q^{\prime 2}$ according to Eq.~(\ref{qil2in}).
\item[QIX1HT:] $x^\prime$ according to Eq.~(\ref{qil2in}).
\item[QIWGT:]  weight (\ref{QIWGT}) associated with the 
               integral (\ref{qiwgt}).
\end{nlist}
\item[\em Procedure:]
\end{nlist}
      The theoretical distributions in $Q^{\prime 2}$ and
      $x^\prime$ are very strongly varying functions. Therefore, the
      improved generation strategy outlined in the description of the
      routine {\bf QIRDIS} is used for default settings of the control
      flags in {\bf QIINIT}.

      The variables $Q^{\prime 2}$ and $x^\prime$ are generated by
      calls to the subroutine {\bf QIRDIS}, which provides  values of 
      \begin{equation}
      \label{qil2in}
      Q^{\prime 2} = (Z_{Q^{\prime 2}})^{-1/N_{Q^{\prime 2}}} {\rm \ 
      and \ }
      x^\prime = (Z_{x^\prime})^{-1/N_{x^\prime}},
      \end{equation}
      with $Z_{Q^{\prime 2}}$ and $Z_{x^\prime}$ 
      uniformly distributed in the intervals 
      \begin{equation}
      \frac{1}{(Q^{\prime 2}_{\rm max})^{N_{Q^{\prime 2}}}}\le
      Z_{Q^{\prime 2}}\le 
      \frac{1}{(Q^{\prime 2}_{\rm min})^{N_{Q^{\prime 2}}}} {\rm \ and \ } 
      \frac{1}{(x^\prime_{\rm max})^{N_{x^\prime}}}\le Z_{x^\prime} \le
      \frac{1}{(x^\prime_{\rm min})^{N_{x^\prime}}}, 
      \end{equation}
      respectively. The optimal powers $N_{Q^{\prime 2}}$ (L2NV) and
      $N_{x^\prime}$ (XPNV) are empirically determined and set in
      {\bf QIHINS}.   

      In analogy to Eq.~(\ref{qirdis-weights}),
      the total weight  $W_{Q^\prime x^\prime\sigma}$, associated with
      the integral (\ref{qiwgt}), 
      is given as a product of three weight factors
      \begin{equation}
      \label{QIWGT}
       W_{Q^\prime x^\prime\sigma} = 
       W_{Q^{\prime 2}}\, W_{x^\prime}\,W_\sigma ,
      \end{equation}
      where 
      \begin{eqnarray}
      \label{AWGT}
      W_{Q^{\prime 2}} &=& \frac{1}{N_{Q^{\prime 2}}}\,
      \left[ (Q^{\prime 2}_{\rm min})^{-N_{Q^{\prime 2}}} -
      (Q^{\prime 2}_{\rm max})^{-N_{Q^{\prime 2}}}\right]\,
      \frac{1}{(Q^{\prime 2})^{-N_{Q^{\prime 2}}}};\\
      \label{BWGT}
      W_{x^\prime} &=& \frac{1}{N_{x^\prime}}\,
      \left[ (x^\prime_{\rm min})^{-N_{x^\prime}} -
      (x^\prime_{\rm max})^{-N_{x^\prime}}\right]\,
      \frac{1}{(x^\prime)^{-N_{x^\prime}}};\\
      \label{qiwgt1}
      W_\sigma &=& 
      \frac{2}{x^\prime}\ 
      Q^{\prime 2}\ \sigma^{(I)}_{q^\prime g}(x^\prime, Q^{\prime 2}).
      \end{eqnarray}  
      The function {\bf Q2SIG} is invoked to calculate the 
      weight factor~(\ref{qiwgt1}) from the instanton-induced 
      $q^\prime g$ total cross section.  

\noindent
SUBROUTINE {\bf QIHPAR}
\begin{nlist}{abcdefghikl}
\item[\em Purpose:]
Generation of the HERWIG~\cite{herwig} identity codes (IDHW) 
of the current quark $k$ as well as of the virtual quark 
$q^\prime$.\\
\item[\em Non-local variables (re)set:]\hfill
\begin{nlist}{abcdefghiklmn}
\item[QINFAM:] = QIPARI(4); number of (light) flavours $n_f$.
\item[QIPINT(1):] IDHW of current quark $k$.
\item[QIPINT(2):] IDHW of virtual quark $q^\prime$.
\item[QICQM2:] $m_{k}^2$; HERWIG mass squared of current quark.
\end{nlist}
\item[{\em Procedure:}]
\end{nlist}
By means of the rejection method, the identity code IDHW of the 
current quark or anti-quark $k$ is chosen randomly
among $2\,n_f$ light quarks and anti-quarks, 
$\{d,u,s,\ldots ;\,\overline{d},\overline{u},\overline{s},\ldots\}$, 
according to the probability distribution
\begin{equation}
      \label{pd2}
      \frac{e_i^2}{\sum\limits_{q=
      d,u,\ldots ;\,\overline{d},\overline{u},\ldots } e_q^2}\,
      ; \hspace{6ex} i=
      d,u,\ldots ;\,\overline{d},\overline{u},\ldots  ,
\end{equation}
reflecting the flavour structure of the standard electromagnetic
$\gamma q_i\overline{q}_i$-coupling. The identity code of the virtual
quark $q^\prime$ is then given in terms of IDHW($k$).
In Eq.~(\ref{pd2}), $e_i$ is the electric charge of the quark flavour $i$.

\noindent
SUBROUTINE {\bf QIINIH}
\begin{nlist}{abcdefghijkl}
\item[\em Purpose:]
Initialization of ``particle'' 'INST'  in HERWIG~\cite{herwig} event record.\\
\item[\em Procedure:]
Association of the name `INST' with the HERWIG 
identity code (IDHW) '206' of the instanton ``particle'' in the 
HERWIG event record.
\end{nlist}

\begin{table}
\begin{center}
{\small
\renewcommand{\arraystretch}{1.05}
\begin{tabular}{|l||c|p{5.5cm}|}\hline
Name & Default value & Description \\ \hline\hline
QIUPAR(1)  &  0.75      & \rule[0ex]{0ex}{3ex}Effective gluon mass\\
QIUPAR(2)  &  $1.0\cdot 10^{-3}$ &$x_{\rm Bj\,min}$:  minimum allowed
$x_{\rm Bj}$\\
QIUPAR(3)  &  1.0 & $z_{\rm max}$: maximum of momentum fraction $z$ 
carried by the gluon\\
QIUPAR(4)  & 1.0  & $y_{\rm Bj\,max}$: maximum allowed $y_{\rm Bj}$\\
QIUPAR(5)  & 0.1 &$y_{\rm Bj\,min}$: minimum allowed $y_{\rm Bj}$ \\
QIUPAR(6)  & 0.9 &$x^\prime_{\rm max}$: maximum allowed $x^\prime$ \\
QIUPAR(7)  & 0.35 &$x^\prime_{\rm min}$: minimum allowed $x^\prime$\\
QIUPAR(8)  & $(\sqrt{\rm QIUPAR(9)}+30)^2$ & $Q^{\prime 2}_{\rm max}$: 
maximum allowed $Q^{\prime 2}$\\
QIUPAR(9)  & $\left(8\,\frac{{\rm QIUPAR}(13)}
{{\rm\bf LAMNF}(3,0.15267)}\right)^2$  & $Q^{\prime 2}_{\rm min}$: 
minimum allowed $Q^{\prime 2}$\\
QIUPAR(10) & QIUPAR(7)  & Cut on $x^\prime$ below which  
the cross section is assumed constant\\
QIUPAR(11) & $-1.0\cdot 10^{-10}$ & Minimum allowed weight for exit of 
main weight generation\\
QIUPAR(12) & 0.0  & $W_{I\,{\rm min}}$: minimum allowed instanton CM energy\\
QIUPAR(13) & {\bf LAMNF}(QIPARI(4),0.219)  & $\Lambda_{\rm
\overline{MS}}^{(n_f)}$:  
from input value (3 loop) 
$\Lambda_{\rm \overline{MS}}^{(5)}=0.219^{+0.025}_{-0.023}$ GeV~\cite{pdg}\\ 
QIUPAR(15) & 0.0  & Minimum energy  after mass subtraction 
from instanton\\
QIUPAR(16) & 0.15 & $\mu_r/Q^\prime$: renorm. scale in
$Q^\prime$ units\\  
QIUPAR(17) & QIUPAR(9) & $Q^2_{\rm min}$: minimum allowed $Q^2$\\
QIUPAR(18) & $\sqrt{{\rm QIUPAR(9)}}$ &
\rule[-2ex]{0ex}{0ex}$\mu_f$: factorization scale\\\hline \hline
QIPARI(1)  & 206  & \rule[0ex]{0ex}{3ex}HERWIG~\cite{herwig} identity code
(IDHW) of instanton  ``particle''\\ 
QIPARI(2)  & 100  & Maximum number of MAMBO iterations per weight\\
QIPARI(3)  & 300  & Maximum number of phase space iterations   
per MAMBO weight\\
QIPARI(4)  & 3 & $n_f$: number of (light) flavours\\
QIPARI(6)  & 10 & Maximum allowed average gluon multiplicity\\
QIPARI(7)  & 20 & Maximum number of iterations in the cross section
weight generation step\\
QIPARI(9)  & 40 & Maximum number of {\bf QIGMUL} iterations\\
QIPARI(10) & 3  & Number of loops in $\alpha_{\overline{\rm MS}}$
evaluation\\
QINWAR(I) & 0 & Counts recurrence of warning I, I~= 1\,..\,14 
(c.\,f. Table~\ref{flags})\\
\hline
\end{tabular}
}
\end{center}
\vspace{3ex}
\caption[dum]{\label{floating} Floating point and integer parameters
set in {\bf QIINIT}. All energy/mass dimensions are in GeV} 
\end{table}

\noindent
SUBROUTINE {\bf QIINIT}
\begin{nlist}{abcdefghikl}
\item[\em Purpose:]
Initialization of input parameters as displayed in Tables~\ref{floating}
and \ref{flags}.\\ 
\item[\em Procedure:]
\end{nlist}
The routine is designed to make the changing of default values as easy as
possible. The parameters available to the user are listed in 
Tables~\ref{floating} and \ref{flags} (for the definition of kinematic
variables see Fig.~\ref{kin-var}). Furthermore, a name for the ``instanton
particle'' is set by QIPARC(1)='INST'. 
After setting the parameters, {\bf QIINIT} calls the subroutine {\bf QIINIH} 
to initialize the new instanton particle in the HERWIG event record. 
Finally, it sets the flag QIRCAL = .TRUE. ($\Leftrightarrow$ calculate derived
quantities and check limits).   

\begin{table}
\begin{center}
{\small
\renewcommand{\arraystretch}{1.1}
\begin{tabular}{|l||c|p{9cm}|}\hline
Name & Default  & Description \\ \hline\hline
QICONT(1) & T & \rule[0ex]{0ex}{3ex}Use non-trivial kinematic weight
for given phase space distribution\\
QICONT(2) & T & Use standard routine (leading-order) to generate
weight for given phase space distribution\\
QICONT(3) & T & Use weight due to instanton cross section\\
QICONT(4) & T &  Disregard instanton minimum mass requirement\\
QICONT(5) & T & Generate $Q^{\prime 2}$ before $x^\prime$\\
QICONT(6) & T & Enforce mass of current quark in kinematics\\
QICONT(9) & T & Enforce maximum allowed number of gluons\\
QICONT(11) & T & Check that energy suffices for gluon generation\\
QICONT(13) & T & Kill events with mass too high\\
QICONT(14) & T & Use $z$ generation as $dz/z$\\
QICONT(15) & T &  Generate $x^\prime$ with efficiency parametrization\\
QICONT(16) & T & Generate $Q^{\prime 2}$ with efficiency parametrization\\
QICONT(18) & T & Use HERWIG rather than 
                 JETSET  hadronization\\
QICONT(20) & T &  Azimuthal angle of $e^\prime$ generated randomly\\
QICONT(21) & T & \rule[-2ex]{0ex}{0ex}Full hadronization on/off\\  
\hline
QIWARF(1)  & F & \rule[0ex]{0ex}{3ex}$Q^{\prime 2}$ upper generation
                 limit reset\\
QIWARF(2)  & F & $Q^{\prime 2}$ lower generation limit reset\\
QIWARF(3)  & F & $x^\prime$ upper generation limit reset\\
QIWARF(4)  & F & $x^\prime$ lower generation limit reset\\
QIWARF(5)  & F & Attempt to (re)set non-existent logical parameter
({\bf QISETF})\\
QIWARF(6)  & F & Attempt to read non-existent logical parameter ({\bf
                 QIGETF})\\ 
QIWARF(7)  & F & Attempt to (re)set non-existent integer parameter
({\bf QISETI})\\ 
QIWARF(8)  & F & Attempt to read non-existent integer parameter ({\bf
                 QIGETI})\\ 
QIWARF(9)  & F & Attempt to (re)set non-existent double precision
parameter ({\bf QISETD})\\
QIWARF(10)  & F & Attempt to read non-existent double precision
                  parameter ({\bf QIGETD})\\
QIWARF(11)  & F & Current quark mass requirement does not fit to the selected 
                  order of variable generation (c.\,f. QICONT(5))\\
QIWARF(12)  & F & Negative $k_T^2$ for current quark, reset to 0.0\\ 
QIWARF(14)  & F & \rule[-2ex]{0ex}{0ex}Too many iterations in 
                 {\bf QIGMUL}\\
\hline
QIRCAL  & T & \rule[0ex]{0ex}{3ex}Check kinematical limits\\
\hline
\end{tabular}
}
\end{center}
\vspace{3ex}
\caption[dum]{\label{flags} Logical flags set in {\bf QIINIT}
(T = .TRUE., F = .FALSE.).} 
\end{table}

\vspace{8ex}
\noindent
SUBROUTINE {\bf QIKGAM}
\begin{nlist}{abcdefghikl}
\item[\em Purpose:]
Generation of the 4-momentum $q$ of the virtual photon,   
with the constraint $-q^2=Sx_{\rm Bj}y_{\rm Bj}\equiv Q^2> 0$.\\
\item[\em Non-local variables  (re)set:]\hfill
\begin{nlist}{abcdefghiklmn}
\item[QIPGAM\,(I):] I-th component of the 4-momentum of 
             the virtual photon, $q=(q_x,q_y,q_z,q_t)$, where
             I=$\{1,2,3\}$ $\Leftrightarrow$ $\{x,y,z\}$ and I=4
             $\Leftrightarrow t$. 
\end{nlist}
\item[\em Procedure:]
\end{nlist}
With the convention $e=(0,0,e_z,e_t)$, $e_z >0$, for the incoming 
lepton momentum in the laboratory frame and assuming $e^2=0$, 
the 4-momentum of the virtual photon is expressed as 
\begin{eqnarray}
\label{photonmomx}
q_x &=&
-\QISUDC \,\cos\XI ,
\\ 
\label{photonmomy}
q_y &=&
-\QISUDC \,\sin\XI ,
\\ 
\label{photonmomz}
q_z &=&
\QISUDA +\frac{S x_{\rm Bj}y_{\rm Bj}}{4\,e_t},
\\
\label{photonmomt}
q_t &=&
\QISUDA -\frac{Sx_{\rm Bj}y_{\rm Bj}}{4\,e_t},
\end{eqnarray}
according to a Sudakov decomposition.
The azimuthal angle $\XI$, $-\pi\leq \XI\leq +\pi$,
is randomly generated for the default setting of the control flag 
QICONT(20) = .TRUE.. Otherwise,
$\XI$ is fixed at zero, i.e. $e^\prime$ (the 4-momentum of the scattered 
$e^\pm$) and $q$ are lying in the $x-z$ plane.

\noindent
SUBROUTINE {\bf QIKGSP}
\begin{nlist}{abcdefghikl}
\item[\em Purpose:]
Generation of the 4-momenta $q^\prime$ and $k$ of the virtual 
quark and the outgoing, on-shell current quark, respectively 
(c.\,f. Fig.~\ref{kin-var}). Implementation of the constraints 
$-q^{\prime 2}=Q^{\prime 2}>0$, where $Q^{\prime 2}$ is the virtuality
generated in {\bf QIHINS}, and 
$k^2=m_{k}^2$.\\  
\item[\em Non-local variables (re)set:]\hfill
\begin{nlist}{abcdefghiklmnop}
\item[QIQGAM\,(I,\,2):] I-th component of the 4-momentum of 
             the virtual quark, $q^\prime
             =(q^\prime_x,q^\prime_y,q^\prime_z,q^\prime_t)$, where 
             I=$\{1,2,3\}$ $\Leftrightarrow$ $\{x,y,z\}$ and I=4
             $\Leftrightarrow t$.  
\item[QIQGAM\,(I,\,1):] I-th component of the 4-momentum of 
             the current quark,
             $k=(k_x,k_y, 
             k_z,k_t)$, where  
             I=$\{1,2,3\}$ $\Leftrightarrow$ $\{x,y,z\}$ and I=4
             $\Leftrightarrow t$.  
\item[QIWARF(12):] = .FALSE. (default); reset to .TRUE. if the 
             discriminant $D^2<0$ in Eq.~(\ref{diskr}).
\item[QINWAR(12):] = 0 (default); counts the number of recurring QIWARF(12) =
                  .TRUE. settings.   
\end{nlist}
\item[\em Procedure:]
\end{nlist}
By convention, the laboratory frame corresponds to the incoming proton
and the incoming lepton travelling in $-z$ and $+z$ direction, respectively.  
Using a Sudakov decomposition to incorporate the required contraints on
$q^{\prime 2}$ and $k^2$, 
\begin{equation}
-q^{\prime 2}=Q^{\prime 2}>0 {\rm \ and\ } k^2=
m_{k}^2,
\end{equation}
the 4-momentum $q^\prime$ of
the virtual quark is expressed as 
\begin{eqnarray}
\label{qprimemomx}
q^\prime_x &=& D\,\cos\PHI -\ALR\,\QISUDC\,\cos\XI
,
\\ \label{qprimemomy}
q^\prime_y &=& D\,\sin\PHI -\ALR\,\QISUDC\,\sin\XI
,
\\ \label{qprimemomz}
q^\prime_z &=& \Op \\ \nonumber &&+ \N - \M
,
\\ \label{qprimemomt}
q^\prime_t &=& -\Op \\ \nonumber &&+ \N + \M 
,
\end{eqnarray}
with
\begin{eqnarray}
\label{diskr}
\lefteqn{
D=} \\ \nonumber && \D \,. 
\end{eqnarray}
The azimuthal angle $\PHI$, $-\pi\leq\PHI\leq +\pi$, is generated randomly.
The 4-momentum of the current quark is then calculated via
\begin{equation}
\label{qprimeprime}
k = q - q^\prime ,
\end{equation}
where the photon 4-momentum $q$ has been generated in {\bf QIKGAM}.

\noindent
SUBROUTINE {\bf QIKPAR}
\begin{nlist}{abcdefghikl}
\item[\em Purpose:]
Calculation of the 4-momentum $g$ of the incoming gluon 
from its generated fractional momentum $z$ and the 4-momentum $P$ of
the incoming proton.\\
\item[\em Non-local variables (re)set:]\hfill
\begin{nlist}{abcdefghiklmn}
\item[QIPPAR\,(I):] I-th component of the 4-momentum of 
             the incoming gluon, $g=(g_x,g_y,g_z,g_t)$, where
             I=$\{1,2,3\}$ $\Leftrightarrow$ $\{x,y,z\}$ and I=4
             $\Leftrightarrow t$. 
\item[QIPPAR\,(5):] $m_g$; HERWIG~\cite{herwig} mass of the incoming gluon.
\end{nlist}
\item[\em Procedure:]
\end{nlist}
The momentum vector $g$ of the incoming, on-shell gluon 
(with effective (HERWIG) mass $\sqrt{g^2}=m_g$) is calculated 
in the laboratory frame according to 
\begin{eqnarray}
\label{glmom}
\lefteqn{
g \equiv (g_x,g_y,g_z,g_t)=}\\ \nonumber
&&\frac{1}{2}\left( 
0,0,z\, (P_t+\mid P_z\mid ) -\frac{m_g^2}{z\, (P_t+\mid P_z\mid )},
z\, (P_t+\mid P_z\mid ) +\frac{m_g^2}{z\, (P_t+\mid P_z\mid )}
 \right) ,
\end{eqnarray}
where $z$ denotes the generated fractional momentum of the incoming gluon and
$P=(0,0,P_z,P_t)$ is the 4-momentum of the incoming proton. If
$P_z\leq 0$,  the sign of $g_z$ is reversed, $g_z\to -g_z$.

\noindent
SUBROUTINE {\bf QIPLST}
\begin{nlist}{abcdefghikl}
\item[\em Purpose:]
Assignment of masses for the $2\,n_f-1 + n_g$ outgoing partons from
the instanton-induced subprocess.\\
\item[\em Non-local variables  (re)set:]\hfill
\begin{nlist}{abcdefghiklmnopqrs}
\item[QIPHEP\,(5,\,J):] mass of the J-th outgoing parton from 
             the instanton-induced subprocess.
\item[QINHEP:] $2\,n_f-1 + n_g$; number of outgoing partons from
             instanton subprocess.
\end{nlist}
\item[\em Procedure:]
\end{nlist}
To each outgoing parton an
appropriate mass (default HERWIG~\cite{herwig} mass for quarks/anti-quarks, 
default gluon mass from {\bf QIINIT}) is assigned.

\noindent
SUBROUTINE {\bf QIPSGN}
\begin{nlist}{abcdefghikl}
\item[\em Purpose:]
Generation of the 4-momenta for the outgoing partons from the
instanton-induced subprocess.\\
\item[\em Non-local variables  (re)set:]\hfill
\begin{nlist}{abcdefghiklmnopqrs}
\item[QIPHEP\,(I,\,J):] I-th component of the 4-momentum corresponding
      to the J-th outgoing parton from the instanton 
      subprocess. Here, I = $\{1,2,3\}$ $\Leftrightarrow$ 
      $\{x,y,z\}$ and I = 4
      $\Leftrightarrow t$. 
\end{nlist}
\item[\em Procedure:]
\end{nlist}
\begin{itemize}
\item By means of the MAMBO~\cite{ks} algorithm\footnote{The MAMBO 
      algorithm implemented in the Monte Carlo generator for baryon
      number violating
      interactions HERBVI~\cite{herbvi} is actually used and provided
      in the source file 'saddle.F' of the QCDINS
      distribution.}, the momenta $p_j$ of the  
      $n=2n_f-1+n_g$ outgoing partons in the $q^\prime g$ CMS are
      generated {\em uniformly} in phase space,
      \begin{equation}
      \label{phasespace}
      \int \prod_{j=1}^{n}
      \left\{ d^4p_j\,\delta^{(+)} \left( p_j^2-m_j^2\right) \right\}
      \, \delta^{(3)}\left( \sum_{k=1}^{n} \vec{p}_k \right)\,
      \delta \left( W_I - \sum_{k=1}^n p_k^t\right) . 
      \end{equation} 
      The masses $m_j$ were assigned in the subroutine {\bf QIPLST}.
\item Next, for the default setting of the control flags in {\bf QIINIT} 
      (QICONT(1) = .TRUE. in Table~\ref{flags}), 
      different energy weights for gluons and quarks  ({\bf QIPSWT})
      are implemented by means of the rejection method. The resulting
      momenta  $p_j$ are then {\em uniformly distributed in energy-weighted
      phase space}, Eq.~(\ref{leadingorder}),  
      corresponding to the leading-order matrix elements~\cite{mrs1}. 
\end{itemize}

\noindent
SUBROUTINE {\bf  QIPSTO}
\begin{nlist}{abcdefghikl}
\item[\em Purpose:]
Store 4-momenta and particle identity codes of all outgoing partons in 
the PHEP common block of HERWIG~\cite{herwig} and set appropriate 
IHEP pointers.\\ 
\item[\em Non-local variables  (re)set:]\hfill
\begin{nlist}{abcdefghiklmnopqrs}
\item[PHEP\,(I,\,IHEP):] I-th 4-momentum component  
             of the J-th outgoing parton from the instanton subprocess, with 
             IHEP=10+J and 
             $1\leq {\rm J}\leq 2n_f-1+n_g$.
\item[IDHEP\,(IHEP):] particle data group~\cite{pdg} identity code (IDPDG) of
             the J-th outgoing parton.
\item[IDHW\,(IHEP):] HERWIG identity code of J-th outgoing parton. 
\item[QIPLIS\,(JLP,\,ILP):] IHEP pointer referring to the
             outgoing parton JLP in the string ILP.
\end{nlist}
\item[\em Remarks:]
See also the descriptions of the routines {\bf QISTID} and {\bf QIGLST}. 
\end{nlist}

\noindent
SUBROUTINE {\bf QIPSWT}\,(PSWGT)
\begin{nlist}{abcdefghijkl}
\item[{\em Purpose:}]
Calculation of the relative energy-weight factor (c.\,f. 
Eq.~(\ref{leadingorder})) for the generated momenta of the outgoing
partons, as corresponding to the
modulus squared of the leading-order matrix elements.\\ 
\item[{\em Argument:}] PSWGT: relative energy weight~(\ref{pswgt}). 
\item[\em Procedure:]
\end{nlist}
An energy-weight factor $w$ corresponding to the modulus squared of the
leading-order matrix elements~\cite{mrs1} 
is calculated as follows\footnote{We thank M. Seymour for pointing out
an error in the previous version of this subroutine and for providing
us with a corrected version~\cite{seymour} of it.}: Each outgoing
quark with four-momentum $p_q$ is weighted by its energy $p_q^t$, each 
outgoing gluon with four-momentum $p_k$ by its energy squared $p_k^{t\,2}$,
such that 
\begin{equation}
\label{kin-weight}
w = \prod_{q=1}^{2n_f-1} p_q^t\ 
    \prod_{k=1}^{n_g} p_k^{t\,2} \, .
\end{equation}
It is easy to show that the corresponding maximum weight is given by 
\begin{equation}
w_{\rm max}=2^{2\,n_g} 
\left[ \frac{W_I}{2\,(n_g+n_f)-1}\right]^{2\,(n_g+n_f)-1} .
\end{equation}
The subroutine {\bf QIPSWT} returns the relative weight 
\begin{equation}
\label{pswgt}
w_{\rm rel} = w/w_{\rm max} 
\end{equation}
in the variable PSWGT.

\noindent
SUBROUTINE {\bf QIPVWT}\,(WT)
\begin{nlist}{abcdefghijkl}
\item[\em Purpose:]
Calculation of the remaining weight factors required for a properly
normalized cross section~(\ref{evwgt}).\\
\item[{\em Argument:}] WT: returned weight factor~(\ref{WT}).
\item[\em Procedure:]
\end{nlist}
The weight factor 
\begin{equation}
\label{WT}
W_n= 2\,\pi\,\alpha^2\, x_{\rm Bj}\, x^\prime \,
\sum\limits_{q^\prime =
          d,u,s,\ldots ;\,
          \overline{d},\overline{u},\overline{s},\ldots }
          e_{q^\prime}^2
\end{equation}
is calculated. 
Here, $\alpha$ is the electro-magnetic fine-structure constant and
$e_{q^\prime}$ denotes the fractional electro-magnetic charge of the
virtual (anti-)quark $q^\prime$. The sum extends over quarks and anti-quarks
of the considered  $n_f$ light  flavours.

\noindent
SUBROUTINE {\bf QIRDIS}\,(LU,\,LL,\,X,\,WGT,\,N)
\begin{nlist}{abcdefghikl}
\item[\em Purpose:]
Strong increase of efficiency through 
generation of $X=Q^{\prime 2},x^\prime$ as $dX/X^{N+1}$.\\ 
\item[\em Arguments:]\hfill
\begin{nlist}{abcdefg}
\item[LU:] logarithm of upper integration limit $X_U$ in Eq.~(\ref{typ-int}).
\item[LL:] logarithm of lower integration limit $X_L$ in Eq.~(\ref{typ-int}).
\item[N:]  power in Eq.~(\ref{zN}).
\item[X:]  $X$ value associated with the uniformly distributed $Z$  
           according to Eq.~(\ref{zN}).
\item[WGT:]Monte Carlo weight~(\ref{qirdis-weights}) with $f\equiv 1$. 
\end{nlist}
\item[{\em Remarks:}]
The instanton-induced cross section~(\ref{evwgt})
involves integrations 
\begin{equation}
\label{typ-int}
I(X_U,X_L)=\int\limits_{X_L}^{X_U} \frac{dX}{X}\,f(X) ,
\end{equation}
where, typically, the function $f(X)$ {\em grows steeply towards small} $X$.
This makes the {\em standard} 
Monte Carlo evaluation (e.\,g. Eqs.~(\ref{yz}), (\ref{Wzg})) of
the integral~(\ref{typ-int}) very inefficient.   
\item[\em Procedure:]
\end{nlist}
The efficiency is strongly increased by means of the present routine.
First, the integration variable is changed to 
\begin{equation}
\label{zN}
Z=X^{-N};\hspace{6ex} dZ=-N\, X^{-(N+1)} dX, 
\end{equation}
with the power $N$ empirically optimized and typically ranging between
2.5 and 5.0. 
Uniformly distributed random points $Z=X^{-N}$ are then generated in
the interval $X_U^{-N}\le Z\le X_L^{-N}$. With this procedure, the
weight corresponding to the integral~(\ref{typ-int}) takes the form
\begin{equation}
\label{qirdis-weights}
W=\frac{1}{N}(X_L^{-N}-X_U^{-N})\,\frac{f(X=Z^{-1/N})}{Z}.
\end{equation} 

\noindent
SUBROUTINE {\bf QISETD}\,(NUM,\,VALUE,\,OK)
\begin{nlist}{abcdefghikl}
\item[\em Purpose:]
Set double precision parameter QIUPAR(NUM) to a desired value.\\  
\item[\em Arguments:]\hfill
\begin{nlist}{abcdefghi}
\item[NUM:] integer, pointing to the parameter QIUPAR(NUM) to be changed.     
\item[VALUE:] desired double precision value of QIUPAR(NUM).      
\item[OK:] logical return flag.    
\end{nlist}
\item[\em Non-local variables (re)set:]\hfill
\begin{nlist}{abcdefghiklmno}
\item[QINWAR(9):] = QINWAR(9)+1; counts attempts to (re)set non-existing
                    parameter QIUPAR(.).
\item[QIWARF(9):] = .TRUE. if parameter index NUM out of range.
\item[QIRCAL:] = .TRUE. if QIWARF(9) = .FALSE. (default).
\end{nlist}
\item[\em Remarks:]
Service routine for resetting the default double precision input
parameters QIUPAR(NUM), NUM=1,2,\ldots,18, set in {\bf QIINIT}
as in Table~\ref{floating}, to the desired 
value VALUE. On return, the logical flag OK is .TRUE. if NUM is within 
its allowed range.  
\end{nlist}

\noindent
SUBROUTINE {\bf QISETF}\,(NUM,\,FLAG,\,OK)
\begin{nlist}{abcdefghikl}
\item[\em Purpose:]
Set logical flag QICONT(NUM) to a desired value.\\
\item[\em Arguments:]\hfill
\begin{nlist}{abcdefghi}
\item[NUM:] integer, pointing to the parameter QICONT(NUM) to be
changed.                   
\item[FLAG:] desired boolean value of QICONT(NUM).       
\item[OK:]  logical return flag. 
\end{nlist}
\item[\em Non-local variables (re)set:]\hfill
\begin{nlist}{abcdefghiklmno}
\item[QINWAR(5):] = QINWAR(5)+1; counts attempts to (re)set non-existing
                    parameter QICONT(.).
\item[QIWARF(5):] = .TRUE. if parameter index NUM out of range.
\item[QIRCAL:] = .TRUE. if QIWARF(5) = .FALSE. (default).
\end{nlist}
\item[\em Remarks:]
Service routine for resetting the default boolean input parameters
QICONT(NUM), NUM=1,2,\ldots,21, set in {\bf QIINIT} as in
Table~\ref{flags}, to the desired value 
FLAG. On return, the logical flag OK is .TRUE. if NUM is within its
allowed range. 
\end{nlist}

\noindent
SUBROUTINE {\bf QISETI}\,(NUM,\,VALUE,\,OK)
\begin{nlist}{abcdefghikl}
\item[\em Purpose:]
Set integer parameter QIPARI(NUM) to a desired value.\\ 
\item[\em Arguments:]\hfill
\begin{nlist}{abcdefghi}
\item[NUM:] integer, pointing to the parameter QIPARI(NUM) to be
changed. 
\item[VALUE:] desired integer value of QIPARI(NUM).           
\item[OK:] logical return flag. 
\end{nlist}
\item[\em Non-local variables (re)set:]\hfill
\begin{nlist}{abcdefghiklmno}
\item[QINWAR(7):] = QINWAR(7)+1; counts attempts to (re)set non-existing
                    parameter QIPARI(.).
\item[QIWARF(7):] = .TRUE. if parameter index NUM out of range.
\item[QIRCAL:] = .TRUE. if QIWARF(7) = .FALSE. (default).
\end{nlist}
\item[\em Remarks:]
Service routine for resetting the default integer input
parameters QIPARI(NUM), NUM=1,2,\ldots,10, set in {\bf QIINIT} as in
Table~\ref{floating}, to the desired  
value VALUE. On return, the logical flag OK is .TRUE. if NUM is within 
its allowed range. 
\end{nlist}

\vspace{8ex}
\noindent
SUBROUTINE {\bf QISPLT}\,(WEIGHT)
\begin{nlist}{abcdefghijkl}
\item[\em Purpose:]
Calculation of the weight associated with the flux of the virtual 
(anti-)quark $q^\prime$ from the $\gamma^\ast q^\prime
k$ vertex (c.\,f. Fig.~\ref{kin-var}).\\ 
\item[{\em Argument:}] WEIGHT: returned weight factor~(\ref{WEIGHT}).
\item[\em Procedure:]
\end{nlist}
The weight  
\begin{equation}
\label{WEIGHT}
W_{q^\prime} = \frac{1}{x^\prime z}\, P_{q^\prime}^{{ (I)}} 
\end{equation}
is calculated, where $P_{q^\prime}^{{ (I)}}$ is the flux
factor~(\ref{flux}) from Refs.~\cite{rs-pl,mrs2}.

\noindent
SUBROUTINE {\bf QISTAT}
\begin{nlist}{abcdefghijkl}
\item[\em Purpose:]
Print input parameter vectors QIUPAR, QIPARI and QICONT, as
well as possible warnings (QIWARF). For default values, see
Tables~\ref{floating} and~\ref{flags}.\\
\end{nlist}

\noindent
SUBROUTINE {\bf QISTID}
\begin{nlist}{abcdefghijkl}
\item[\em Purpose:]
Generation of identity codes, 4-momenta and colour/flavour 
connections of the $2\,n_f -1+n_g$ outgoing partons of the instanton-induced
subprocess $g+q^\prime \stackrel{I}{\Rightarrow} (2\,n_f-1)\cdot
q+n_g\cdot g$.\\  
\item[\em Procedure:]
\end{nlist}
\begin{itemize}
\item Given are the identity codes of the incoming partons, the number
      $n_g+1$ of gluons and the required ``flavour democracy'' of the 
      possible partonic subprocesses like e.\,g. for $n_f=3$ and
      $n_g=4$, 
      \begin{equation}
      \label{proc-ex}
      s+g\stackrel{I}{\Rightarrow} d + \overline{d} + u + \overline{u} + 
      s + 4\cdot g  \, .
      \end{equation}
      The structure as well as the identity codes of the partonic final state 
      are then set up by the following two steps:\\
\begin{itemize}
\item[$\circ$] First, $n_f$ $[\,q \ldots \overline{q}\,]$ - ``strings'' of
      partons are generated ({\bf QIGLST}). Each begins with a quark, 
      followed by a random number $\le n_g+1$ of gluons, and ends with
      an anti-quark of randomly chosen flavour. Due to flavour
      democracy, $q\overline{q}$-pairs of all $n_f$ flavours must
      occur precisely once, e.\,g. for the process~(\ref{proc-ex}) above,
      \begin{equation}
      \label{strings}
      [\,d\; g\; g\;\overline{s}\,]\hspace{5ex} [\,u\;g\;\overline{d}\,]
      \hspace{5ex} 
      [\,s\;g\;g\;\overline{u}\,]\, .
      \end{equation} 
      All partons are marked as outgoing at this stage.   
\item[$\circ$] Next ({\bf QIGPAR}), one of the $n_g+1$ gluons in the
      above strings is randomly marked as incoming. Furthermore, the 
      anti-parton corresponding to the virtual (anti-)quark $q^\prime$
      is identified and marked as incoming, e.\,g. ($\bullet$) for the
      process~(\ref{proc-ex}) and string configuration~(\ref{strings}) 
      above, 
      \begin{equation}
      \label{example3}
      [\,d\; g\; g\;\stackrel{\bullet}{\textstyle\overline{s}}\,]\hspace{5ex} 
      [\,u\;g\;\overline{d}\,]
      \hspace{5ex} 
      [\,s\;g\;\stackrel{\bullet}{\textstyle{g}}\;\overline{u}\,]\, .
      \end{equation} 
      \hfill\\
\end{itemize}
\item The assignment of the final state parton momenta is based 
      on the crucial requirement of isotropy in the ``instanton rest frame'', 
      $\vec{q}^{\,\prime} +\vec{g}=\vec{0}$. It proceeds through the
      following steps:\\ 
\begin{itemize}       
\item[$\circ$] The appropriate (HERWIG~\cite{herwig}) mass $m_i$ is
      assigned to each of the $2n_f-1+n_g$ outgoing partons~$i$ ({\bf
      QIPLST}).   
\item[$\circ$] The 4-momenta $p_i$ in the instanton rest frame
      are generated uniformly in    
      energy-weighted phase-space~(\ref{leadingorder}),
      corresponding to the leading-order matrix element~\cite{mrs1} 
      with different energy weights for gluons and quarks ({\bf QIPSGN}).
\item[$\circ$] The four-momenta and particle identity codes of all outgoing 
      partons are stored in the PHEP common block of HERWIG 
      ({\bf QIPSTO}). The corresponding IHEP pointers start at IHEP=11
      for the first outgoing parton from the instanton subprocess (a
      $d$ quark in our example~(\ref{example3}) above) and are
      increased consecutively along the strings, e.\,g. for the
      process~(\ref{proc-ex}) and string
      configuration~(\ref{example3}) above, the IHEP pointers would read  
      \begin{center}
      $$\begin{array}{lccccrclcccrclccccl}
      [ & d\ & g\ & g\ &
      \stackrel{\bullet}{\textstyle\overline{s}}\ & ]\hspace{5ex} & & 
      [ & u\ & g\ & \overline{d}\ & ]\hspace{5ex} & &      
      [ & s\ & g\ & \stackrel{\bullet}{\textstyle{g}}\ &
      \overline{u}\ & ] 
      \\ 
          & 11\  & 12\  & 13\  &     \                 &     & &
          & 14\  & 15\  & 16\  & \    & &
          & 17\  & 18\  &   \  & 19\ &     .
      \end{array}$$
      \end{center}
\hfill\\ 
\end{itemize}
\item Finally ({\bf QICCON}), the stored list of IHEP pointers is
      augmented by those of the current quark (IHEP=10) and of the incoming
      gluon (IHEP=6), as assigned in {\bf QIHGEN}. For the example
      above, this means
      \begin{center}
      $$\begin{array}{lccccrclcccrclccccl}
      [ & d\ & g\ & g\ &
      \stackrel{\bullet}{\textstyle\overline{s}}\ & ]\hspace{5ex} & & 
      [ & u\ & g\ & \overline{d}\ & ]\hspace{5ex} & &      
      [ & s\ & g\ & \stackrel{\bullet}{\textstyle{g}}\ &
      \overline{u}\ & ] 
      \\ 
          & 11\  & 12\  & 13\  & {\bf 10}    \                 &     & &
          & 14\  & 15\  & 16\  & \    & &
          & 17\  & 18\  & {\bf 6}  \  & 19\ &     .
      \end{array}$$
      \end{center}
      \hfill\\
      \noindent
       As a main issue, the colour/flavour connections are established 
       in {\bf QICCON} as follows:
      The colour flow is constructed by
      connecting the colour lines of 
      neighbouring partons within each of the strings. The flavour
      flow is constructed by connecting the flavour lines of the 
      quark at the beginning of a string with the flavour line of 
      the anti-quark at the end of a string.   
      For the example above, the corresponding HERWIG
      mother/daughter pointers specifying the colour/flavour flow are
      then as in Table~\ref{jmohep}. 
      \begin{table} 
      \begin{center}
{\small
\renewcommand{\arraystretch}{1.2}
      \begin{tabular}{|r|r|c|c|}\hline
      IHEP & Description & JMOHEP(2,IHEP) & JDAHEP(2,IHEP)\\ \hline
      11   & $d$ out     & 12   & 10\\
      12   & $g$ out & 13  &  11\\
      13   & $g$ out     & 10  & 12\\
      10   & $\overline{s}$ out (current quark) &  11 & 13 \\ \hline
      14   & $u$ out     & 15  & 16\\
      15   & $g$ out     & 16  & 14\\
      16   & $\overline{d}$ out     & 14  & 15\\ \hline
      17   & $s$ out     & 18  & 19\\
      18   & $g$ out & 6  & 17\\
      6    & $g$ in      & 19  & 18 \\
      19   & $\overline{u}$ out & 17 & 6  \\
      \hline
      \end{tabular}
      }
      \vspace{2ex}
      \caption[dum]{\label{jmohep}HERWIG~\cite{herwig}
      mother/daughter pointers specifying the colour/flavour flow for
      the example~(\ref{proc-ex}) in the description of subroutine
      {\bf QISTID}.} 
      \end{center}
      \end{table}
\end{itemize}

\noindent
SUBROUTINE {\bf QIUSPS\,(WT)}
\begin{nlist}{abcdefghikl}
\item[{\em Purpose:}] Dummy routine for optional modification of
        relative phase space weight as calculated in {\bf QIPSWT}.\\
\item[{\em Arguments:}] WT: modified relative phase space weight.
\item[{\em Remarks:}] Called from {\bf QIPSWT} if QICONT(2) =
                 .FALSE.. Calculation of WT has to be provided by the user.
\end{nlist}

\noindent
FUNCTION {\bf XI}\,
(XPR,\,XI\_MIN,\,XI\_MAX,\,XQ,\,XMU,\,NF,\,LOOPFL)
\begin{nlist}{abcdefghikl}
\item[\em Purpose:]
The saddle point value of the conformally invariant $\iai$-separation
$\xi$ is calculated as function of $\xpr$, 
$4\pi/\alpha_{\overline{\rm MS}}(Q^\prime )$, 
$4\pi/\alpha_{\overline{\rm MS}}(\mu_r )$ and $n_f$.\\
\item[\em Arguments:]\hfill\\
\begin{nlist}{abcdefghi}
\item[XPR:] $\xpr$
\item[XI\_MIN:]  $\xi_{\rm min}$; lower boundary of $\xi$ used for 
interpolation.
\item[XI\_MAX:]  $\xi_{\rm max}$; upper boundary of $\xi$ used for 
interpolation. 
\item[XQ:] $4\pi/\alpha_{\overline{\rm MS}}(Q^\prime )$
\item[XMU:] $4\pi/\alpha_{\overline{\rm MS}}(\mu_r )$       
\item[NF:]  $n_f$; number of light flavours.      
\item[LOOPFL:]
\begin{nlist}{abcd}
\item[= 1:] 1-loop renormalization group (RG) invariance~\cite{rs-pl} along 
            with 1-loop form of $\alpha_s$.
\item[= 2:] 2-loop RG invariance~\cite{rs-pl} along with
            2-loop form of $\alpha_{\overline{\rm MS}}$.
\item[= 3:] (default) 2-loop RG invariance along with
            3-loop form of $\alpha_{\overline{\rm MS}}$.
\end{nlist}
\end{nlist}
\item[\em Procedure:]
\end{nlist}
First, the saddle point equation~(\ref{eq2}) is solved analytically for
$x^\prime$, and the $x^\prime_i$-values corresponding to a discrete set
of $\xi_i$ values are calculated as  
\begin{equation}
\label{xprast}
x^\prime_i = \frac{(\xi_i -2)}
{(\xi_i +2)+4 \ts(\xi_i) (\ts(\xi_i)-2 v_i)/v_i^2} ,
\end{equation}
with $\xi_{\rm min}\le\xi_i\le \xi_{\rm max}$ and $v_i$ from
Eqs.~(\ref{eq1})--(\ref{dadada}) inserted:  
\begin{eqnarray}
\label{vastxq}
\lefteqn{
v_i=v_i (\xi_i, X(Q^\prime ), X(\mu_r )) =2\, \dts\,\times }
\\[2.5ex] \nonumber && \hspace{-5ex}
W\left( 
\frac{
\exp\left\{
\frac{1}{2}\frac{\ts}{\dts}
+\frac{1}{2} \frac{\Delta_1-1}{\Delta_1\beta_0}
\left[ \Delta_1 \ln
\left( \frac{\Delta_1 X(\mu_r )}{X(Q^\prime )+(\Delta_1-1)X(\mu_r )}\right)
-1
\right] X(\mu_r )
+\frac{1}{2}\frac{X(Q^\prime )}{\beta_0}
\right\}}
{2\,\dts}
\right) .
\end{eqnarray} 
In Eq.~(\ref{vastxq}), the explicit dependence of $v_i$ on $Q^\prime/\mu_r$ 
(c.\,f. Eq.~(\ref{eq1}) has been eliminated in favour of an
$X(Q^\prime )$, $X(\mu_r )$ dependence by means of the standard scale
transformation of $\alpha_{\overline{\rm MS}}$. 
Here, we have introduced the shorthand
\begin{equation}
X(M )\equiv \frac{4\pi}
{\alpha_{\overline{\rm MS}}\left(M \right)}.
\end{equation}

The desired continuous inversion $\xi_\ast=\xi_\ast(\xpr,\ldots)$ is
then achieved by means of numerical interpolation based on the above exact
supporting points ($\xi_i,x^\prime_i$). 

\noindent
FUNCTION {\bf XQS}\,(QLAM,\,LOOPFL,\,NF)
\begin{nlist}{abcdefghikl}
\item[{\em Purpose:}]
Calculation of $4\pi/
\alpha_{\overline{\rm MS}}$ as function of
$M/\Lambda_{\overline{\rm MS}}^{(n_f)}$, where $M$ is a generic mass scale.\\ 
\item[\em Arguments:]\hfill
\begin{nlist}{abcdefghik}
\item[QLAM:] $M/\Lambda_{\overline{\rm MS}}^{(n_f)}$
\item[LOOPFL:]
\begin{nlist}{abcd}
\item[= 1:] 1-loop form of $\alpha_s$.
\item[= 2:] 2-loop form of $\alpha_{\overline{\rm MS}}$.
\item[= 3:] (default) 3-loop form of $\alpha_{\overline{\rm MS}}$.
\end{nlist}      
\item[NF:]  $n_f$; number of light flavours.        
\end{nlist}
\vspace{8ex}
\item[\em Procedure:]
\end{nlist}
The running coupling $\alpha_{\overline{\rm MS}}$
is calculated according to 
the explicit formula Eq.~(9.5a) in Ref.~\cite{pdg}, which is accurate
to 3-loop. The quantity $4\pi/\alpha_{\overline{\rm MS}}$ is returned.  
The integer variable LOOPFL = $1,\,2,\,3$ controls the loop-order at which
Eq.~(9.5a) in  Ref.~\cite{pdg} is truncated.  

\section{\label{usage}Usage and availabiliy}

QCDINS 2.0 should be loaded together with HERWIG~\cite{herwig} version 5.9
and JETSET~\cite{jetset} version 7.4, that are part of the CERNLIB 
distribution. 
The program is a slave system, which the user must call from his own 
steering program. A very simple example is provided in Appendix A. 
By default, QCDINS is compiled in form of a library, 
'libqcdins.a', that may be linked together with 'libherwig59.a',
'libjetset74.a' and the library 'libpdflib.a' of parton distribution
functions to the steering program. An extensive demonstration 
program, including an interface to the HZTOOL~\cite{hztool} package, and a 
detailed installation instruction are included in the distribution. 

Information about the distribution, its update history, an interactive
manual, the source code, pictures of typical events etc. can be
accessed via the QCDINS WWW site,
http://www.desy.de/\~{}t00fri/qcdins/qcdins.html.

\begin{ack}
First of all, we are grateful to M. Gibbs who left physics in May 1995. 
Without his early contributions the QCDINS package would presumably not exist.
A number of people helped improving the code. In particular, let us mention
T. Carli, G. Grindhammer, H. Jung and M. Seymour. Specifically, we thank
H. Jung for his efforts to incorporate an interface to the JETSET package.
Moreover, we are grateful to M. Seymour for an evaluation of the QCDINS package
during the DESY Workshop 1998/99 on Monte Carlo Generators for HERA
Physics. Finally, we thank G. Ingelman for a careful reading of the
manuscript.   
\end{ack}

\section*{Appendix A: Simple steering program}

The following code may serve as the simplest illustration of a
steering program for the QCDINS/HERWIG package. Note that the two
subroutines {\bf HWANAL} and {\bf HWAEND} must exist.

For an example of an interface with the JETSET hadronization, we refer
to the extensive steering program 'qtesthz' included in the QCDINS
distribution. 

{\scriptsize
\begin{verbatim}
      PROGRAM QCDINS
#include "herwig.inc"
C Force inclusion of block data
      EXTERNAL HWUDAT

C Initialize process number
      IPROC = 17600
C Maximal number of events in this run
      MAXEV = 1000
C Beam particles 
      PBEAM1 = 27.5D0
      PBEAM2 = 820.0D0
      PART1  = 'E+'
      PART2  = 'P'
C Initialize common blocks
      CALL HWIGIN
C User can reset parameters at this point, otherwise values
C set in HWIGIN will be used.
C No vertex information in event printout 
      PRVTX=.FALSE.
C Reset number of shots for initial max weight search
      IBSH = 5000
      LRSUD=0
      LWSUD=77
C Seeds 
      NRN(1)=106645412
      NRN(2)=135135176
C Use laboratory frame
      USECMF = .FALSE.
C Compute parameter-dependent constants
      CALL HWUINC
C Number of HERWIG events to print out
      MAXPR = 1
C Call hwusta to make any particle stable
      CALL HWUSTA('PI0     ')
      CALL HWUSTA('K_S0    ')

C Initialize default QCDINS input parameters
      CALL QIINIT
C Print input parameters 
      CALL QISTAT
C Loop over events
      CALL QCLOOP

C User's terminal calculations
      CALL HWAEND
      STOP
      END

      SUBROUTINE HWANAL
      RETURN
      END

      SUBROUTINE HWAEND
      RETURN
      END
\end{verbatim}
}

\section*{Appendix B: Test run output}

Below, we display the essential output from a test run  
of the very simple steering program from Appendix A.
This test run simulates 1000 complete instanton-induced events
in deep-inelastic $e^+P$ scattering (HERA), with $E_{e^+}=27.5$ GeV and 
$E_P=820$ GeV, in the laboratory frame. 
All the QCDINS parameters correspond
to the default values as set in the initialization routine {\bf QIINIT}
(c.\,f. Tables~\ref{floating} and \ref{flags}).     

For reasons of space, the display of the  first event has been
truncated after the hard subprocess level.
 
\vspace{3ex}
  
{\scriptsize
\hspace{35ex}\ldots

\hspace{35ex}\ldots
\begin{verbatim}
      INPUT CONDITIONS FOR THIS RUN

      BEAM 1 (E+      ) MOM. =     27.50
      BEAM 2 (P       ) MOM. =    820.00
      PROCESS CODE (IPROC)   =   17600
\end{verbatim}
\hspace{35ex}\ldots

\hspace{35ex}\ldots
\begin{verbatim}
 =============================================================
 QCD Instanton Monte Carlo Information             Version 2.0
 =============================================================
 Parameter                                             Value  
 =============================================================
 Default gluon mass (GeV)                           0.7500E+00
 Minimum allowed value of x_bj                      0.1000E-02
 Maximum allowed value of z (proton mom.frac.)      0.1000E+01
 Maximum allowed value of y_bj                      0.1000E+01
 Minimum allowed value of y_bj                      0.1000E+00
 Maximum allowed value of x prime                   0.9000E+00
 Minimum allowed value of x prime                   0.3500E+00
 Maximum allowed value of Q prime **2 (GeV**2)      0.1652E+04
 Minimum allowed value of Q prime **2 (GeV**2)      0.1134E+03
 Lower cut for ME calculation on x prime            0.3500E+00
 Lowest allowed weight efficiency cut               -.1000E-09
 Minimum instanton invariant mass                   0.0000E+00
 Lambda-MS-bar(nf) [GeV] from PDG 1998              0.3459E+00
 Minimum total K.E. of outgoing partons (GeV)       0.0000E+00
 Renormalization point KAPPA = mu_r/Qprime:         0.1500E+00
 Minimum allowed value of Q **2 (GeV**2)            0.1134E+03
 Factorization scale (GeV)                          0.1065E+02
 =============================================================
 ID code assumed for instanton                             206
 Maximum number of MAMBO iterations per PS wt.             100
 Maximum number of PS wt. rejections per event             300
 Number nf of (light) flavours                               3
 Maximum average gluon multiplicity in distbn.              10
 Maximum number of iterations for ME generation             20
 Maximum number of QIGMUL iterations                        40
 Number of loops in RG-invariance/alpha_s                    3
 =============================================================
 Control flag option                                  Setting 
 =============================================================
 Reweight phase space configurations                      True
 Use default phase space reweighting                      True
 Use Matrix element weight                                True
 Disregard instanton minimum mass requirement             True
 Generate Q prime before x prime                          True
 Enforce mass of current quark in kinematics              True
 Enforce limit on maximum number of gluons                True
 Ensure mass less than subprocess energy                  True
 Kill events with insufficient instanton mass             True
 Use z generation as dz/z                                 True
 Use x prime **-n generation for efficiency               True
 Use Q prime **-n generation for efficiency               True
 Use HERWIG rather than JETSET hadronization              True
 Use random azimuth angle for scattered electron          True
 Use full hadronization                                   True
 =============================================================

          INITIAL SEARCH FOR MAX WEIGHT

          PROCESS CODE IPROC =       17600
          RANDOM NO. SEED 1  =     1246579
                     SEED 2  =     8447766
          NUMBER OF SHOTS    =        5000
   
   
                ****************************
                *                          *
                *    QCDINS version 2.0    *
                *                          *
                ****************************
\end{verbatim}
\hspace{35ex}\ldots

\hspace{35ex}\ldots
\begin{verbatim}
          NEW MAXIMUM WEIGHT =  2.4520378923431534E-05
          NEW MAXIMUM WEIGHT =  6.0858601240747906E-02
          NEW MAXIMUM WEIGHT =  0.1289870599803515    
          NEW MAXIMUM WEIGHT =  0.3662942637479250    
          NEW MAXIMUM WEIGHT =  0.5705336816482935    
          NEW MAXIMUM WEIGHT =  0.8061467165013312    

          INITIAL SEARCH FINISHED

          OUTPUT ON ELEMENTARY PROCESS

          NUMBER OF EVENTS   =           0
          NUMBER OF WEIGHTS  =        5000
          MEAN VALUE OF WGT  =  2.8721E-02
          RMS SPREAD IN WGT  =  6.4377E-02
          ACTUAL MAX WEIGHT  =  7.3286E-01
          ASSUMED MAX WEIGHT =  8.0615E-01

          PROCESS CODE IPROC =       17600
          CROSS SECTION (PB) =   28.72    
          ERROR IN C-S  (PB) =  0.9104    
          EFFICIENCY PERCENT =   3.563    



EVENT      1:   27.50 GEV/C E+       ON  820.00 GEV/C P         PROCESS: 17600

SEEDS:  106645412 &  135135176   STATUS: 100  ERROR:   0  WEIGHT: 0.2872E-01
\end{verbatim}
\vspace{8ex}
\begin{verbatim}
                          ---INITIAL STATE---    

IHEP    ID    IDPDG IST MO1 MO2 DA1 DA2    P-X     P-Y     P-Z   ENERGY   MASS 
  1 E+          -11 101   0   0   4   0    0.00    0.00   27.50   27.50    0.00
  2 P          2212 102   0   0   0   0    0.00    0.00 -820.00  820.00    0.94
  3 CMF           0 103   1   2   0   0    0.00    0.00 -792.50  847.50  300.33

                          ---INITIAL STATE---    

IHEP    ID    IDPDG IST MO1 MO2 DA1 DA2    P-X     P-Y     P-Z   ENERGY   MASS 
  4 E+          -11 101   1   0   0   0    0.00    0.00   27.50   27.50    0.00

                         ---HARD SUBPROCESS---   

IHEP    ID    IDPDG IST MO1 MO2 DA1 DA2    P-X     P-Y     P-Z   ENERGY   MASS 
  5 E+          -11 121   7   9  21   9    0.00    0.00   27.50   27.50    0.00
  6 GLUON        21 122   7  20  22  19    0.00    0.00  -50.37   50.38    0.75
  7 HARD          0 120   5   6   9  20    0.53    0.99  -22.98   77.98   74.51
  8 INST          0   3   7   0   0   0    0.23    9.33  -43.54   47.01   15.07
  9 E+          -11 123   7   5  26   5    8.20   -7.47    3.48   11.63    0.00
 10 UBAR         -2 124   7  14  27  17   -8.43   -1.86   17.19   19.24    0.31
 11 DQRK          1 124   7  12  31  13   -0.07    0.52   -7.65    7.67    0.32
 12 GLUON        21 124   7  13  33  11    0.16   -0.28   -0.64    1.04    0.75
 13 DBAR         -1 124   7  11  35  12   -0.67   -0.12   -2.65    2.75    0.32
 14 UQRK          2 124   7  15  37  10    0.01    4.97  -11.73   12.75    0.32
 15 GLUON        21 124   7  16  39  14    0.53    0.62   -2.17    2.44    0.75
 16 GLUON        21 124   7  17  41  15    0.43   -0.43   -8.44    8.49    0.75
 17 GLUON        21 124   7  10  43  16   -0.72    0.09   -2.40    2.62    0.75
 18 SQRK          3 124   7  19  45  20   -0.11    2.50   -4.93    5.55    0.50
 19 GLUON        21 124   7   6  47  18    0.81    0.93   -1.46    2.06    0.75
 20 SBAR         -3 124   7  18  51   6   -0.14    0.52   -1.47    1.64    0.50
 21 Z0/GAMA*     23   3   5   7   0   0   -8.20    7.47   24.02   15.87  -21.17
\end{verbatim}
\hspace{35ex}\ldots

\hspace{35ex}\ldots
\begin{verbatim}
 CHECK OF ENERGY-MOMENTUM CONSERVATION IN THE EVENT:
 ==============================================================
  
 Sum_i(P_i)[GeV] = (  4.87E-13 -5.49E-13  1.59E-12 -4.20E-12 )
  
 with  60 stable particles in final state contributing.

          OUTPUT ON ELEMENTARY PROCESS

          NUMBER OF EVENTS   =        1000
          NUMBER OF WEIGHTS  =       27258
          MEAN VALUE OF WGT  =  2.8837E-02
          RMS SPREAD IN WGT  =  6.3371E-02
          ACTUAL MAX WEIGHT  =  7.0199E-01
          ASSUMED MAX WEIGHT =  8.0615E-01

          PROCESS CODE IPROC =       17600
          CROSS SECTION (PB) =   28.84    
          ERROR IN C-S  (PB) =  0.3838    
          EFFICIENCY PERCENT =   3.577    
\end{verbatim}
}
\vfill\eject

\end{document}